\documentclass[a4paper,11pt]{article}
\usepackage{cite}
\usepackage{epsfig}
\usepackage{amssymb,amsmath}
\usepackage{graphicx}
\usepackage{float}
\usepackage{comment}
\usepackage{epstopdf}
\usepackage{slashed}
\usepackage{caption3} % load caption package kernel first
\DeclareCaptionOption{parskip}[]{} % disable "parskip" caption option
\usepackage[small]{caption}
\usepackage[utf8]{inputenc}
\allowdisplaybreaks
\usepackage[section]{placeins}
\usepackage{float}

\pdfoutput=1

\newlength{\dinwidth}
\newlength{\dinmargin}
\newcommand{\cR}{\mathcal{R}}

\newcommand{\bel}[1]{\be\label{#1}}

\newcommand{\ta}[0]{\tilde \alpha}

\def\be{\begin{equation}}
\def\ee{\end{equation}}
\def\beqn{\begin{eqnarray}}
\def\eeqn{\end{eqnarray}}

\def\ba{\begin{array}{c}}
\def\bat{\begin{array}{cc}}
\def\ea{\end{array}}
\def\bi{\begin{itemize}}
\def\ei{\end{itemize}}

\def\be{\begin{equation}}
\def\ee{\end{equation}}

\begin{document}

\title{
\begin{flushright}\vbox{\normalsize \mbox{}\vskip -6cm
FTUV/15-0213 \\[-3pt] IFIC/15-09
}
\end{flushright}\vskip 45pt
{\bf New Barr-Zee contributions to $\mathbf{(g-2)_\mu}$ in two-Higgs-doublet models}}
\bigskip

\author{Victor Ilisie \\[15pt]
{\small IFIC, Universitat de Val\`encia -- CSIC, Apt. Correus 22085, E-46071 Val\`encia, Spain}
\date{}
}

\maketitle
\bigskip \bigskip
%\vspace{-1.cm}

\begin{abstract}

We study the contribution of new sets of two-loop Barr-Zee type diagrams to the anomalous magnetic moment of the muon within the two-Higgs-doublet model framework. We show that some of these contributions can be quite sizeable for a large region of the parameter space and can significantly reduce, and in some cases even explain, the discrepancy between the theoretical prediction and the experimentally measured value of this observable. Analytical expressions are given for all the calculations performed in this work.

\end{abstract}

\newpage

\section{Introduction}
\label{sec:introd}

Now that a SM-like Higgs particle has been experimentally discovered \cite{Aad:2012tfa,Aad:2013wqa,Chatrchyan:2012ufa,Chatrchyan:2013lba,Aaltonen:2012qt}, the possibility of an enlarged scalar sector becomes very plausible. In this analysis we are going to use the anomalous magnetic moment of the muon as a probe for new physics and study new contributions to this observable within the two-Higgs-doublet model (2HDM) framework.
The anomalous magnetic moment of the muon has been extensively analysed within the Standard Model (SM) and its numerous extensions.
Even if the SM prediction still suffers from large theoretical uncertainties (mostly hadronic and electroweak) it is a nice place to look for new physics. The latest result for the discrepancy between the SM prediction and the experimental measured value is given
by \cite{Broggio:2014mna,Wang:2014sda,Aoyama:2012wk,Czarnecki:1995wq,Czarnecki:1995sz,Gnendiger:2013pva,
Jegerlehner:2009ry,Davier:2010nc,Krause:1996rf,Prades:2009tw,Colangelo:2014dfa,Colangelo:2014pva,
Blum:2014oka,Pauk:2014rfa,Kurz:2014wya,Colangelo:2014qya,Blum:2013xva,Melnikov:2006sr,Davier:2004gb,
Passera:2004bj,Knecht:2003kc,Bennett:2006fi,Agashe:2014kda}
\begin{align}
\Delta a_\mu^{exp} \; \equiv \; a_\mu^{exp} \; - \; a_\mu^{\text{SM}} \; = \; 262 (85) \, \times \, 10^{-11} \; .
\label{deltaamuexp}
\end{align}
Here we will study the two-loop Barr-Zee type \cite{Barr:1990vd} contributions to $\Delta a_\mu$ that have not been analysed previously within the  2HDM. We show that some of these diagrams can bring rather sizeable contributions for a quite large region of the parameter space and therefore can reduce the value of the difference between theory and experiment given by (\ref{deltaamuexp}). We also show that other sets of these type of diagrams bring small contributions and can be safely discarded. For the calculations we use the most generic Higgs potential and the generic Yukawa structure of the aligned two-Higgs-doublet model (A2HDM) \cite{Pich:2009sp}. Thus, we also
re-examine the classical Barr-Zee type diagrams \cite{Broggio:2014mna,Wang:2014sda,Dedes:2001nx,Gunion:2008dg,Chang:2000ii,Cheung:2001hz,Krawczyk:2002df,Larios:2001ma,Cheung:2003pw,
Arhrib:2001xx,Heinemeyer:2003dq,Kong:2004um,Cheung:2009fc,Bian:2014zka} expressing their contributions in terms of the three independent complex alignment parameters $\varsigma_{u,d,l}$.    
All the results are given in analytical form. The phenomenological analysis is made assuming a CP-conserving Lagrangian. However, all the generic formulae given in this work can be used for future, and more complete, analyses without assuming CP-conservation. Additional constraints coming from the flavour sector and global fits to the LHC data are also taken into account  
\cite{ilisie3,ilisie2,ilisie1,ilisie00,ilisie01,JungTuzon,PichTuzon1,PichDstar,PichEDP,PichBll}.

In the first part of this paper, section~\ref{sec:A2HDM}, we present the relevant features of the A2HDM. In section~\ref{sec:oneloop} we present the one-loop results in terms of the generic A2HDM parameters. In section~\ref{sec:twoloop} we present the classical two-loop Barr-Zee results and the calculation of the new sets of this type of diagrams that can potentially bring sizeable contributions to the anomalous magnetic moment.
Section~\ref{sec:phenomeno} is dedicated to the phenomenological analysis for the CP-conserving case and the presentation of the relevant contributions. Finally, we conclude in section \ref{sec:conclusions} with a brief summary of our results. One appendix is also given, with technical details for the calculation of a particular set of Barr-Zee type diagrams.

%%%%%%%%%%%%%%%%%%%%%%%%%%%%%%%%%%%%%%%%%%%%%%%%%%%%%%%%%%%%%%%%%%%%%%%%%%%%%%%%%%%%%%%%%%%%%%%%%%%%%%%%%%%%%%%%%%%%%%%%%%
%%%%%%%%%%%%%%%%%%%%%%%%%%%%%%%%%%%%%%%%%%%%%%%%%%%%%%%%%%%%%%%%%%%%%%%%%%%%%%%%%%%%%%%%%%%%%%%%%%%%%%%%%%%%%%%%%%%%%%%%%%
%%%%%%%%%%%%%%%%%%%%%%%%%%%%%%%%%%%%%%%%%%%%%%%%%%%%%%%%%%%%%%%%%%%%%%%%%%%%%%%%%%%%%%%%%%%%%%%%%%%%%%%%%%%%%%%%%%%%%%%%%%

\section{The Aligned Two-Higgs-Doublet Model}
\label{sec:A2HDM}

The 2HDM extends the SM with a second scalar doublet of hypercharge $Y=\frac{1}{2}$.
It is convenient to work in the so-called Higgs basis
$(\Phi_1,\Phi_2)$, where only one doublet acquires a vacuum expectation value:
\begin{equation}  \label{Higgsbasis}
\Phi_1=\left[ \begin{array}{c} G^+ \\ \frac{1}{\sqrt{2}}\, (v+S_1+iG^0) \end{array} \right] \; ,
\qquad\qquad\qquad
\Phi_2 = \left[ \begin{array}{c} H^+ \\ \frac{1}{\sqrt{2}}\, (S_2+iS_3)   \end{array}\right] \; ,
\end{equation}
where $G^\pm$ and $G^0$ denote the Goldstone fields.
Thus, $\Phi_1$ plays the role of the SM scalar doublet with
$v = (\sqrt{2}\, G_F)^{-1/2} = 246~\mathrm{GeV}$.
The physical scalar spectrum contains five degrees of freedom: two charged fields $H^\pm(x)$
and three neutral scalars $\varphi_i^0(x)=\{h(x),H(x),A(x)\}$, which are related with the $S_i$ fields
through an orthogonal transformation $\varphi^0_i(x)=\mathcal{R}_{ij} S_j(x)$.
The form of the $\mathcal{R}$ matrix is fixed by the scalar potential, which determines the neutral scalar mass matrix
and the corresponding mass eigenstates. A detailed discussion is given in \cite{ilisie3,ilisie2,ilisie1}. In general, the CP-odd component $S_3$ mixes with the CP-even fields
$S_{1,2}$ and the resulting mass eigenstates do not have a definite CP quantum number.
If the scalar potential is CP symmetric this admixture disappears; in this particular case, $A(x) = S_3(x)$
and
\bel{eq:CPC_mixing}
\left(\ba h\\ H\ea\right)\; = \;
%\left[\bat \cos{(\alpha - \beta)} & \sin{(\alpha - \beta)} \\ -\sin{(\alpha - \beta)} & \cos{(\alpha - \beta)}\ea\right]\;
\left[\bat \cos{\tilde\alpha} & \sin{\tilde\alpha} \\ -\sin{\tilde\alpha} & \cos{\tilde\alpha}\ea\right]\;
\left(\ba S_1\\ S_2\ea\right) \, .
\ee
Performing a phase redefinition of the neutral CP-even fields, we can fix the sign of $\sin{\ta}$.  In this work we adopt the conventions\ $M_h \le M_H$\ and\
$ 0 \leq \ta \leq \pi$, so that $\sin{\ta}$ is positive.

The most generic Yukawa Lagrangian with the SM fermionic content gives rise to FCNCs because the fermionic couplings of the two scalar doublets cannot be simultaneously diagonalized in flavour space. The non-diagonal neutral couplings can be eliminated by requiring the alignment in flavour space of the Yukawa matrices~\cite{Pich:2009sp}; {\it i.e.}, the two Yukawa matrices coupling to a given type of right-handed fermions are assumed to be proportional to each other and can, therefore, be diagonalized simultaneously. The three proportionality parameters $\varsigma_f$~($f=u,d,l$) are arbitrary complex numbers and introduce new sources of CP violation.

In terms of the fermion mass-eigenstate fields, the Yukawa interactions of the A2HDM read~\cite{Pich:2009sp}
\beqn\label{lagrangian}
 \mathcal L_Y & = &  - \frac{\sqrt{2}}{v}\; H^+ \left\{ \bar{u} \left[ \varsigma_d\, V M_d \mathcal P_R - \varsigma_u\, M_u^\dagger V \mathcal P_L \right]  d\, + \, \varsigma_l\, \bar{\nu} M_l \mathcal P_R l \right\}
\nonumber \\
& & -\,\frac{1}{v}\; \sum_{\varphi^0_i, f}\, y^{\varphi^0_i}_f\, \varphi^0_i  \; \left[\bar{f}\,  M_f \mathcal P_R  f\right]
\;  + \;\mathrm{h.c.} \, ,
\eeqn
where $\mathcal P_{R,L}\equiv \frac{1\pm \gamma_5}{2}$ are the right-handed and left-handed chirality projectors,
$M_f$ the diagonal fermion mass matrices
and the  couplings of the neutral scalar fields are given by:
\begin{equation}    \label{yukascal}
y_{d,l}^{\varphi^0_i} = \cR_{i1} + (\cR_{i2} + i\,\cR_{i3})\,\varsigma_{d,l}  \, ,
\qquad\qquad
y_u^{\varphi^0_i} = \cR_{i1} + (\cR_{i2} -i\,\cR_{i3}) \,\varsigma_{u}^* \, .
\end{equation}
The usual models with natural flavour conservation, based on discrete ${\cal Z}_2$ symmetries, are recovered for particular (real) values of the couplings $\varsigma_f$ \cite{Pich:2009sp}. The coupling of a single neutral scalar with a pair of gauge bosons takes the form ($V=W,Z$)
\begin{align}
g_{\varphi_i^0 VV} = \mathcal{R}_{i1} \; g^{\text{SM}}_{hVV}\, ,
\label{sumrule}
\end{align}
which implies $g_{hVV}^2 + g_{HVV}^2 + g_{AVV}^2 = (g_{hVV}^\text{SM})^2$. Thus, the strength of the SM Higgs interaction is shared by the three 2HDM neutral bosons. In the CP-conserving limit, the CP-odd field decouples while the strength of the $h$ and $H$ interactions is governed by the corresponding $\cos\tilde\alpha$ and $\sin\tilde\alpha$ factors. Again, for further details about the interaction Lagrangian as well as the Higgs potential, needed for the calculations in this work, see \cite{ilisie3,ilisie2,ilisie1}.

%%%%%%%%%%%%%%%%%%%%%%%%%%%%%%%%%%%%%%%%%%%%%%%%%%%%%%%%%%%%%%%%%%%%%%%%%%%%%%%%%%%%%%%%%%%%%%%%%%%%%%%%%%%%%%%%%%%%%%%%%%
%%%%%%%%%%%%%%%%%%%%%%%%%%%%%%%%%%%%%%%%%%%%%%%%%%%%%%%%%%%%%%%%%%%%%%%%%%%%%%%%%%%%%%%%%%%%%%%%%%%%%%%%%%%%%%%%%%%%%%%%%%
%%%%%%%%%%%%%%%%%%%%%%%%%%%%%%%%%%%%%%%%%%%%%%%%%%%%%%%%%%%%%%%%%%%%%%%%%%%%%%%%%%%%%%%%%%%%%%%%%%%%%%%%%%%%%%%%%%%%%%%%%%

\section{One-loop contribution}
\label{sec:oneloop}

\begin{figure}[!htb]
\centering
\includegraphics[scale=0.45]{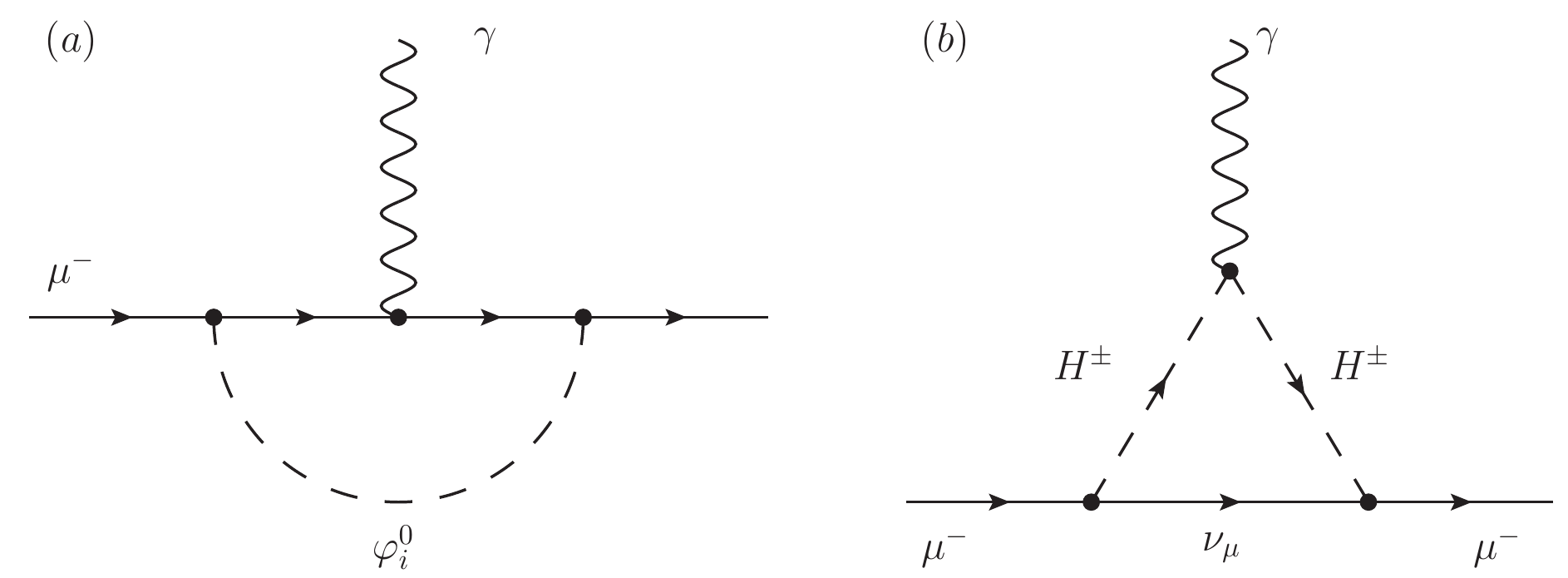}
\caption{\it{One-loop contribution to $\Delta a_\mu$ in two-Higgs-doublet models.}}
\label{oneloop1}
\end{figure}

\noindent At the one-loop level, the contribution of the 2HDM extension of the SM to the anomalous magnetic moment of the muon is given by the two well-known diagrams shown in Fig.~\ref{oneloop1}. The explicit expressions for these contributions, in terms of the most generic Higgs potential and the A2HDM Yukawa structure, are given by 
\begin{align}
\Delta a_\mu^{(a)} = \frac{m_\mu^2}{8 \pi^2 v^2}  \, \sum_i \, \frac{m_\mu^2}{M_{\varphi_i^0}^2}  \!\ \Bigg[ &  \!\  \text{Re}\big( y^{\varphi_i^0}_l \big)^2  \int_0^1 dx  \!\ \frac{x^2(2-x)}{(m_\mu^2/M_{\varphi_i^0}^2)x^2-x+1}  \notag \\ 
&  \, \, + \, 
\text{Im}\big( y^{\varphi_i^0}_l \big)^2  \int_0^1 dx  \!\ \frac{-x^3}{(m_\mu^2/M_{\varphi_i^0}^2)x^2-x+1}  \!\ 
\Bigg] \; ,
\end{align}
for the neutral Higgses and
\begin{align}
\Delta a_\mu^{(b)} = \frac{m_\mu^2}{8 \pi^2 v^2}  \!\ \Bigg(\frac{m_\mu^2}{M_{H^\pm}^2}\Bigg)  \!\  |\varsigma_l|^2  \int_0^1 dx  \!\ \frac{x^2(1-x)}{(m_\mu^2/M_{H^\pm}^2)x(1-x)-x} \; ,
\end{align}
for the charged Higgs. These contributions have been previously analysed in \cite{Haber:1978jt,Leveille:1977rc,Krawczyk:1996sm,Queiroz:2014zfa,Broggio:2014mna,Dedes:2001nx,Larios:2001ma}.

It's a known fact that the two-loop Bar-Zee type diagrams dominate over the one-loop contributions. The two loop contributions have a loop suppression factor of $(\alpha/\pi)$ but also have an enhancement factor of $(M^2/m_\mu^2)$, where $M$ stands for the mass of heavy particles running in one of the loops: $M_{H^\pm}$, $m_{t}$, $M_{\varphi_i^0}$, etc. This last factor usually dominates over the first one. Furthermore, in the usual $\mathcal{Z}_2$ models, there is an extra enhancement (suppression) factor from $\tan\beta$ ($\cot\beta$) for some diagrams. In the aligned model there is a lot more freedom to independently enhance or suppress any contribution through the alignment parameters $\varsigma_f$. We shall see next, that for somewhat large values of these parameters, there are new Barr-Zee contributions that have never been taken into account, and can bring quite sizeable contributions to $(g-2)_\mu$.

\section{Two-loop contribution}
\label{sec:twoloop}

\begin{figure}[!htb]
\centering
\includegraphics[scale=0.55]{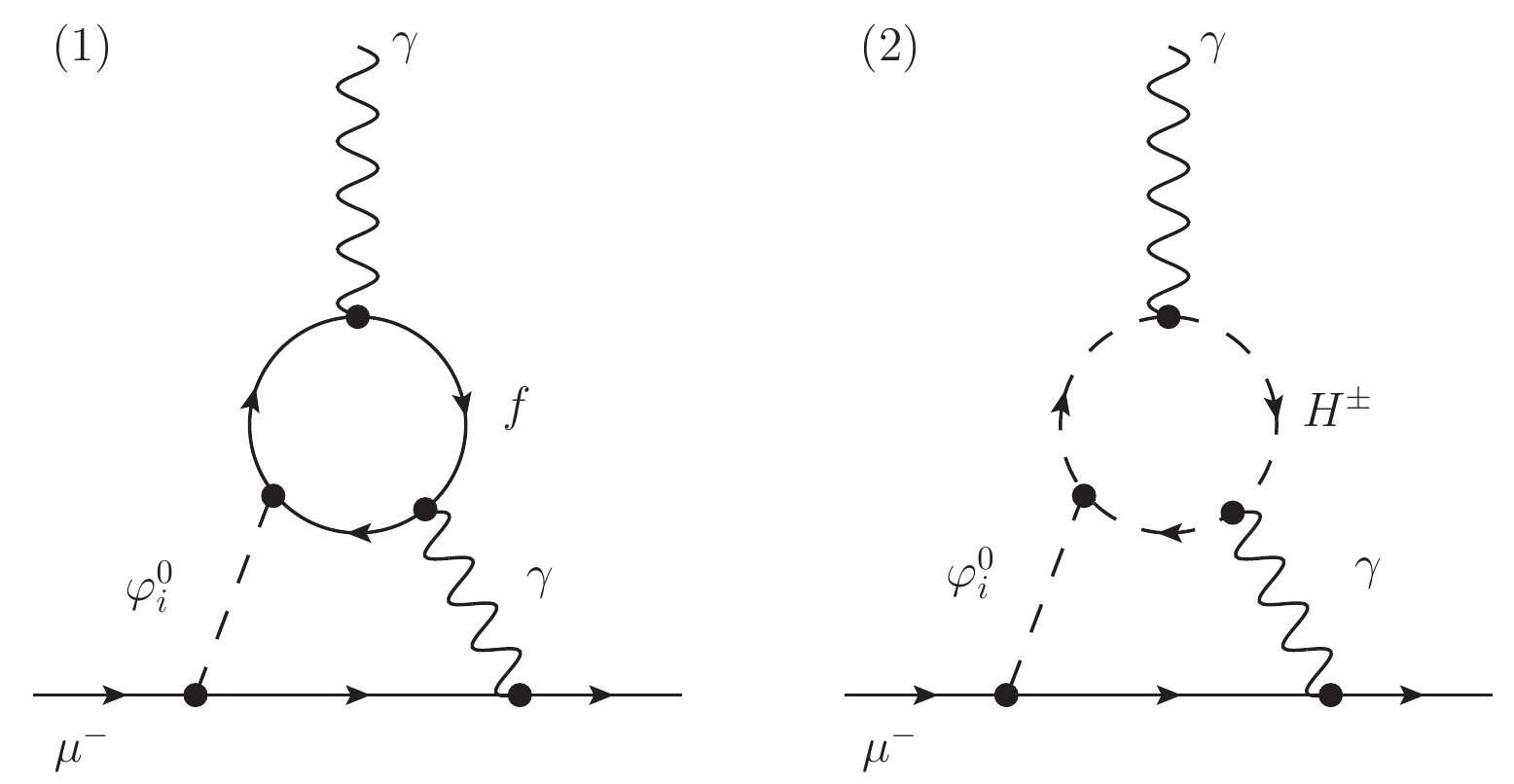} $ \!\  \!\ $
\includegraphics[scale=0.55]{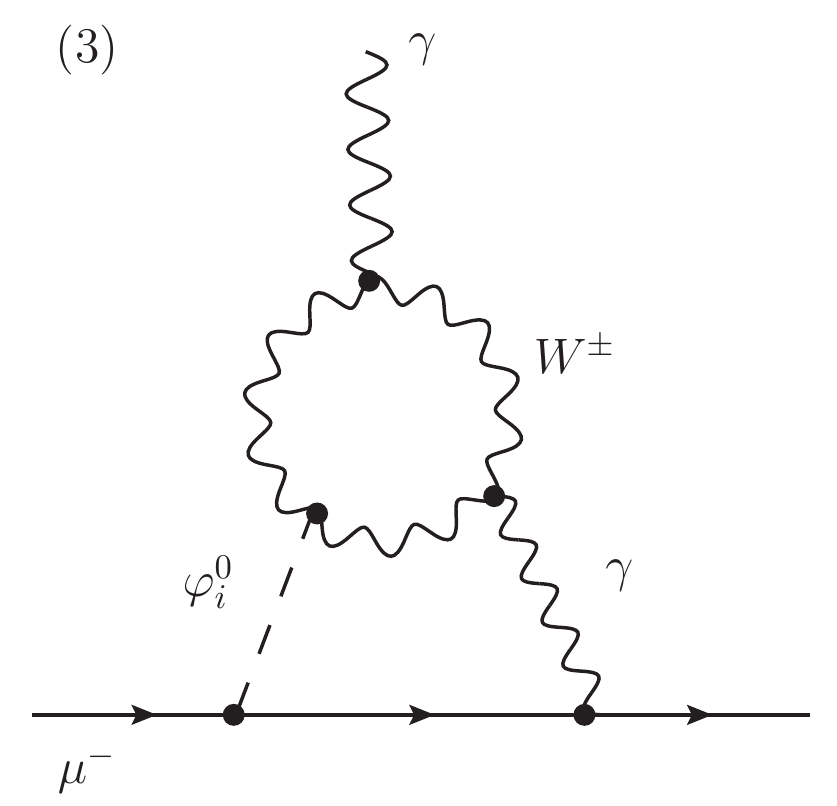}
\caption{{\it Two-loop Barr-Zee type (with an internal photon) contribution to $\Delta a_\mu$ in two-Higgs-doublet models .}}
\label{BarZee1}
\end{figure}

The Barr-Zee type contributions with an internal photon, {\it i.e.}, Fig.~\ref{BarZee1}, diagrams (1) and (2), have been extensively analysed within the 2HDM and also in minimal super-symmetry (MSSM) framework  \cite{Broggio:2014mna,Wang:2014sda,Dedes:2001nx,Gunion:2008dg,Chang:2000ii,Cheung:2001hz,Krawczyk:2002df,Larios:2001ma,Cheung:2003pw,
Arhrib:2001xx,Heinemeyer:2003dq,Kong:2004um,Cheung:2009fc,Bian:2014zka}. Diagram (3) from Fig.~\ref{BarZee1} is also of the Barr-Zee type and could, in principle bring important contributions. Given that the coupling to a pair of gauge bosons of the recently discovered scalar particle is close to the SM prediction \cite{ilisie3}, one expects the contributions from the remaining scalars to be somewhat suppressed (by a factor $\mathcal{R}_{i1}$). However, we shall see that this statement is not correct, and that this contribution is quite sizeable.

Similar contributions to the ones shown in Fig.~\ref{BarZee1}, but with the internal photon replaced by a Z boson have been also analysed in the literature \cite{Chang:2000ii}. These contributions have a relative suppression factor of order $10^{-2}$. This factor is in part due to the vectorial couplings of $Z$ to leptons, which are the only ones that survive for both scalar and pseudo-scalar bosons \cite{Chang:2000ii}, and in part from the Z propagator which introduces a new mass scale $M_Z$. Therefore we will ignore these contributions in our present analysis.

\begin{figure}[!htb]
\centering
\includegraphics[scale=0.55]{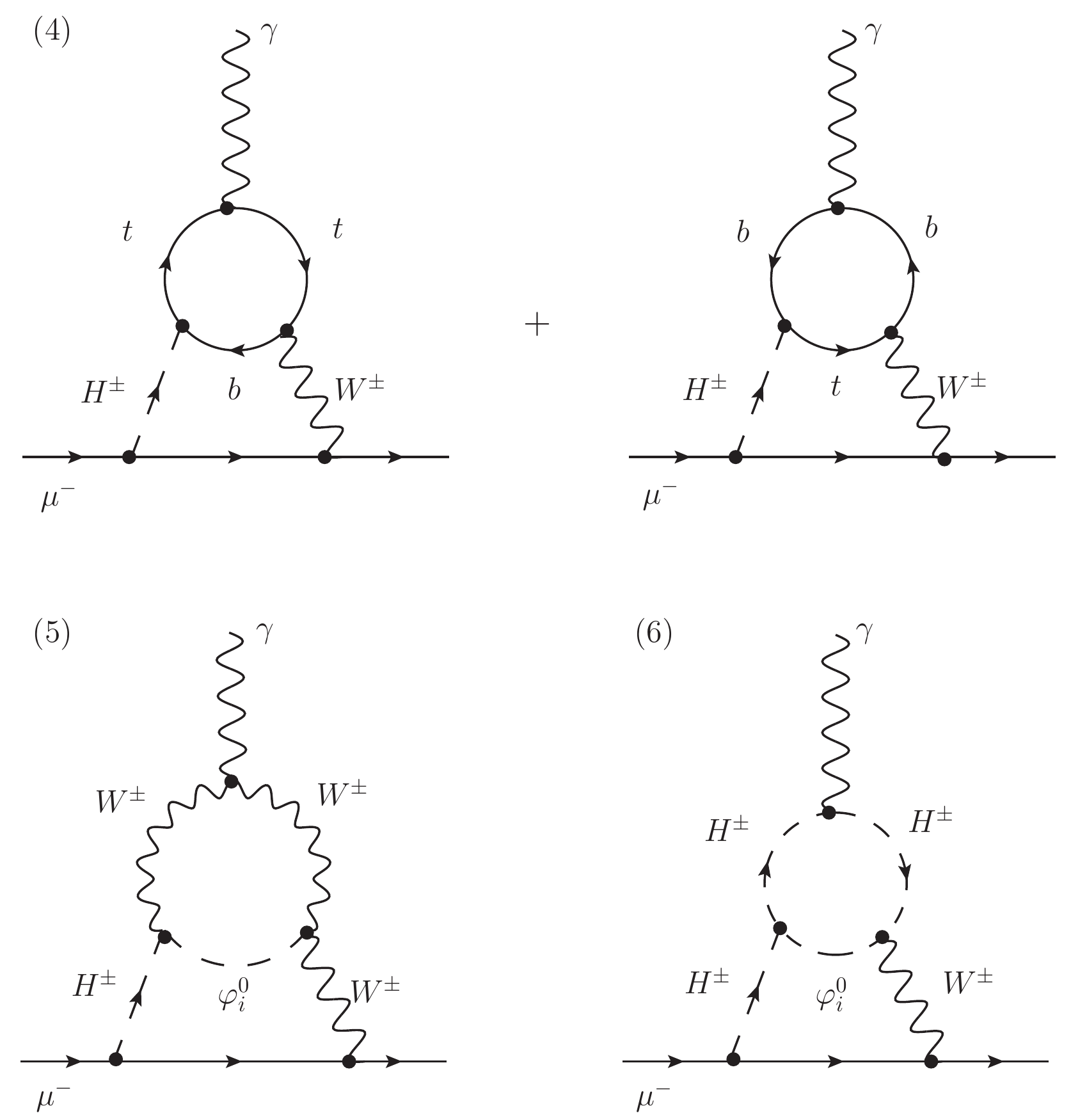}
\caption{{\it Two-loop Barr-Zee type (with a charged Higgs and an internal W boson) contribution to $\Delta a_\mu$ in two-Higgs-doublet models.}}
\label{BarZee2}
\end{figure}

\begin{figure}[!htb]
\centering
\includegraphics[scale=0.55]{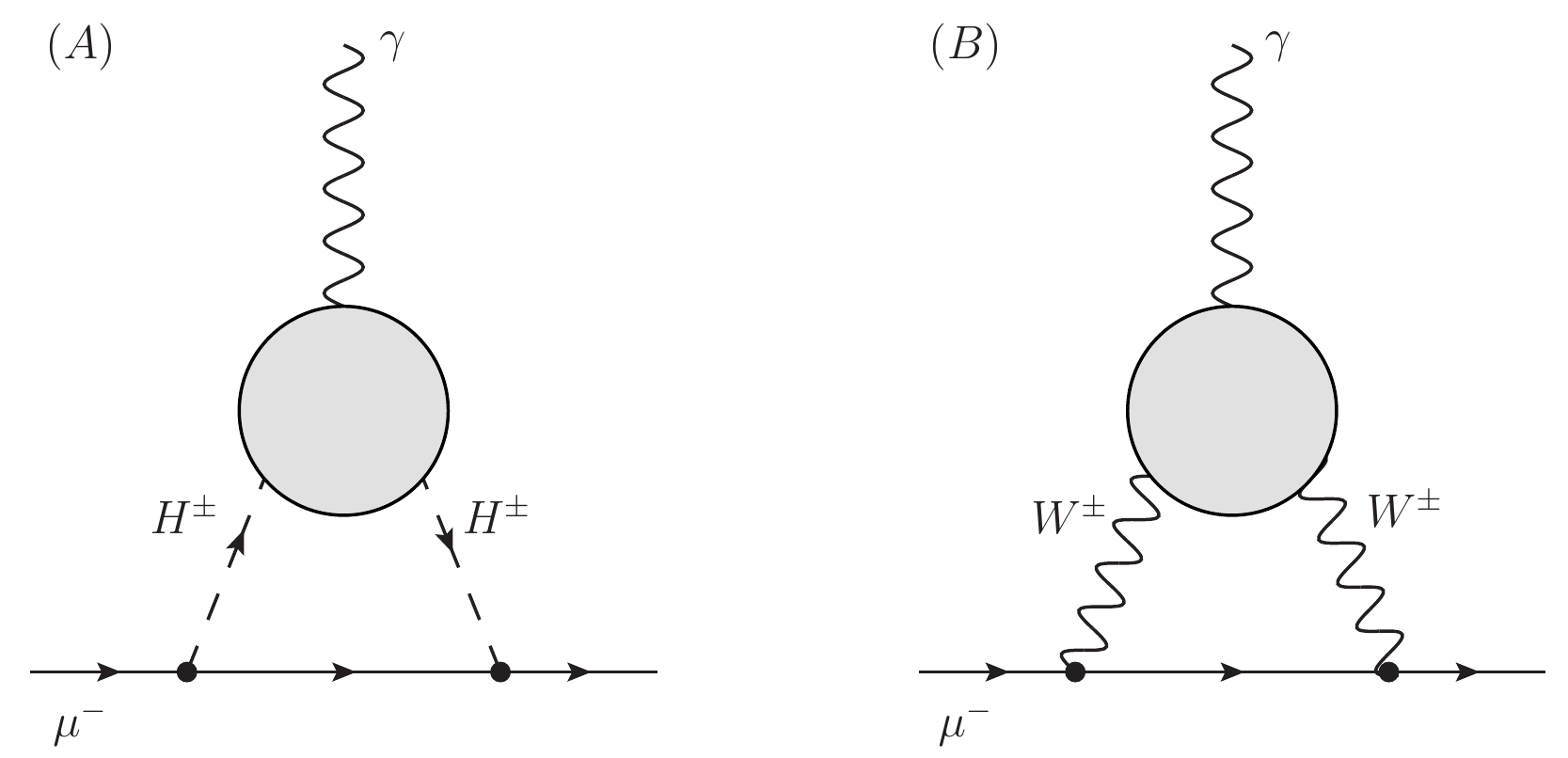}
\caption{{\it Generic two-loop Barr-Zee type contributions, with two internal charged Higges (left) and two internal W bosons (right), to $\Delta a_\mu$ in two-Higgs-doublet models.}}
\label{BarZeeAB}
\end{figure}

This is, pretty much, the summary of all the mechanisms that are usually considered in the literature. However, there is no reason a priori to discard other similar Barr-Zee contributions with a charged Higgs $H^\pm$ substituting the neutral scalars $\varphi_i^0$, and a $W$ boson substituting the internal photon\footnote{Similar contributions, however, with sfermionic loops within the MSSM framework have been previously analysed in \cite{Chen:2001kn}.}. These diagrams are illustrated in Fig.~\ref{BarZee2}. On one hand, one expects a relative suppression factor with respect to the contributions of the diagrams from Fig.~\ref{BarZee1} due to the propagator of the $W$ boson (note that in this case we don't have the additional suppression factor due to the gauge boson couplings to leptons, as in the $Z$ case). On the other hand, one must also expect to be able to re-enhance these contributions with the $\varsigma_f$ (or $\tan\beta$) parameters, and therefore, obtain sizeable contributions at least in some regions of the parameter space.

In this analysis we shall calculate the contribution from these new diagrams and demonstrate, that in fact, all of these new sets can bring rather sizeable contributions to the anomalous magnetic moment of the muon in a quite large region of the parameter space. For completeness we shall also present the classical two-loop results in terms of the most generic Higgs potential and in terms of the generic Yukawa texture of the A2HDM.

Before moving on to the next section and presenting the analysis, there are a couple of related cases that are worth discussing. They are shown in Fig.~\ref{BarZeeAB}, where the grey circles stand for the same loop contributions as in Fig.~\ref{BarZee2} (excluding the fermionic loops for diagram (B) which is just a pure SM contribution). The contribution from the first case (A), will have a relative suppression factor $m_\mu^2/M_W^2$ with respect to the contributions of diagrams from Fig.~\ref{BarZee2} so we can safely discard it. The contribution coming from the second set, Fig.~\ref{BarZeeAB} (B), does not have this suppression factor, thus we can expect, at least in principle, a rather sizeable effect. Details of the the full calculation of this last set of diagrams, together with other technical details are given in appendix \ref{appendixa}. Roughly one obtains a contribution of $\mathcal{O}(10^{-11})$ which is rather small and we shall not include it in this analysis. 

Next we move on to the analysis of the set of diagrams shown in Fig.~\ref{BarZee2} which is the main goal of our paper.

\subsection{Gauge invariant effective vertices}

\begin{figure}[]
\centering
\includegraphics[scale=0.55]{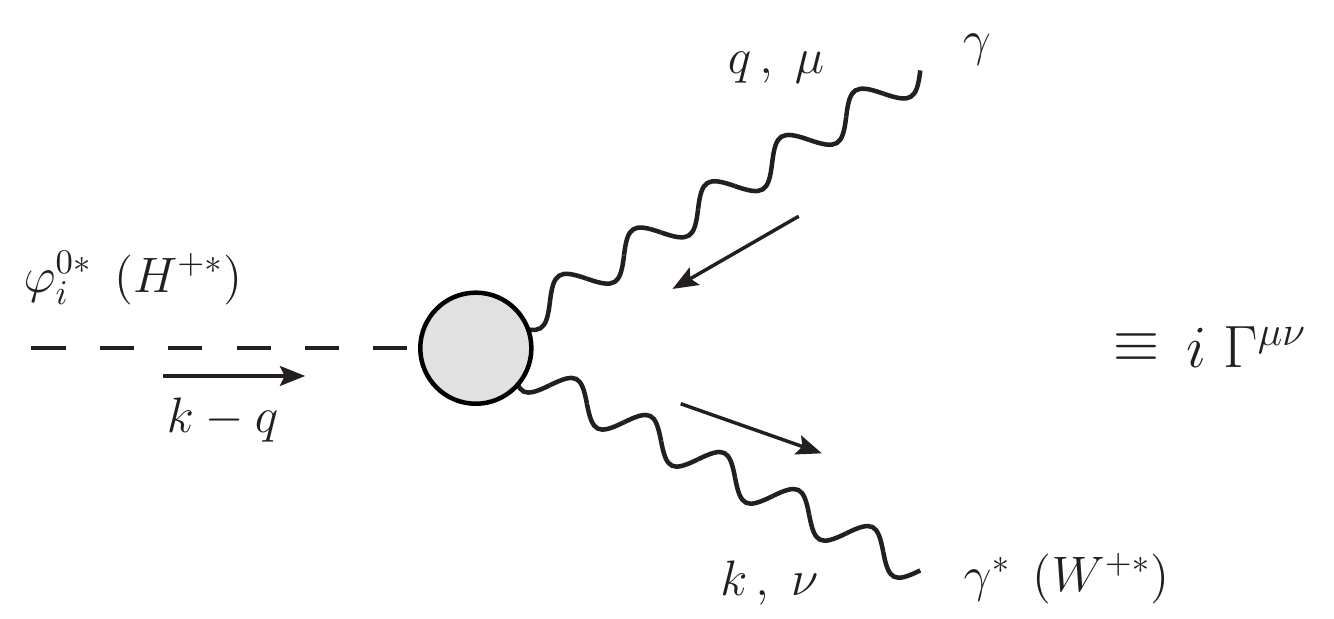}
\caption{{\it Feynman rule for the gauge-invariant one loop effective vertices $\varphi_i^{0}-\gamma\gamma$ and $H^{+} - W^{+}\gamma$.}}
\label{gamamunu}
\end{figure}

The calculation of the two-loop Barr-Zee type diagrams can be separated in two parts. We will first calculate the $\varphi_i^{0}-\gamma\gamma$ and $H^{+} -\gamma W^{+}$ one-loop effective vertices and obtain analytical and rather simple expressions. With these expressions, the calculation of the second loop becomes quite trivial. The effective vertices can be written in a generic gauge-invariant transverse form:
\begin{align}
i \!\  \Gamma^{\mu \nu} = i \!\  (g^{\mu\nu} k\cdot q - k^\mu q^\nu) \!\  S  \!\ +  \!\ i \!\  \epsilon^{\mu\nu\alpha\beta} \!\  k_\alpha \!\  q_\beta \!\  \tilde{S}  \; ,
\label{transverse}
\end{align}
where $q^\mu$ is the momentum of the incoming real photon and $k^\nu$ is the momentum of the out-going virtual gauge boson 
(see Fig.~\ref{gamamunu}), and where $S$ and $\tilde{S}$ are scalar form factors.  In order to obtain this expression we have considered the most generic Lorentz structure for the $\Gamma^{\mu\nu}$ vertex, and we have imposed the electromagnetic current conservation $q_\mu \, \Gamma^{\mu\nu}=0$. All terms proportional to $q^\mu$ have also been eliminated as they cancel when contracted with the polarization vector of the photon. As the $W$ boson is off-shell, in the actual calculation of the effective vertex there will also appear some other Lorentz structures than the ones shown in (\ref{transverse}). However in some cases, these gauge-dependent contributions vanish when calculating the second loop or they are cancelled by some other non Barr-Zee terms, as it is nicely shown in \cite{Abe:2013qla}. If this was not the case, when summing the proper non Barr-Zee contributions to the gauge dependent Barr-Zee terms, the result must be gauge independent. As the gauge dependence from the Barr-Zee terms is cancelled by other sub-dominant topologies, we also expect this contribution to be sub-dominant. Therefore, we shall discard these terms in our analysis. 

The gauge independent contribution from each set represented by the generic topologies in Fig.~\ref{BarZee1} and Fig.~\ref{BarZee2} is transverse by itself, {\it i.e.,} of the form given in (\ref{transverse}); we can therefore decompose the results into eight separate contributions.
For the $\varphi_i^{0}-\gamma\gamma$ effective vertex
$S=S_{(1)} + S_{(2)} + S_{(3)}$ and $\tilde{S}= \tilde{S}_{(1)}$; as for the $H^{+}-\gamma W^{+}$ vertex we have 
$S=S_{(4)} + S_{(5)} + S_{(6)}$ and $\tilde{S}= \tilde{S}_{(6)}$. Note that the only contributions to the $\epsilon^{\mu\nu\alpha\beta} \, k_\alpha \, q_\beta$ structure come from the fermionic loops. Furthermore, one can adopt our strategy from \cite{ilisie1}, and further simplify the calculations of $S_{(j)}$ by only considering the terms that contribute to the structure $k^\mu \, q^\nu$. 

It is worth mentioning the following technical detail. When performing the calculations for the first loop, after introducing the Feynman parametrization and after integrating over the four-momentum, one obtains a denominator similar to   
\begin{align}
[k^2 x(x-1) + M_a^2 x + M_b^2(1-x) + k \cdot q \, 2y \,x(1-x)]^{-1} \, ,
\end{align}
where $M_{a,b}$ are the masses of $heavy$ particles running in the loop, {\it i.e.,} $M_W, \, m_t, \, M_{H^\pm}\, , $ etc. It is a very common assumption that the photon is ``soft" so one can ignore the $k \cdot q$ term as a good approximation. This term, in fact, can be safely ignored without making any assumptions on the ``softness'' of the photon. Keeping track of this term, one can observe that it simply vanishes when calculating the second loop integral. However, this happens accidentally for diagrams (1) to (6); for the $WW\gamma$ effective vertices calculated in appendix~\ref{appendixa}, this is not always the case. Thus, having checked that these terms play no role in our present case, we will discard them already at the one-loop level in order to give simpler and more elegant expressions for the form factors $S_{(i)}$ and $\tilde S_{(i)}$. After performing the four-momentum loop integral we obtain the following expressions for the scalar form-functions
\begin{align}
S_{(1)} & \, = \, \sum_{i,f} \, \frac{\alpha \, m_f^2}{\pi \, v} \, Q_f^2  \, N_C^f \,\, \text{Re}\big( y^{\varphi_i^0}_f \big) \, \int_0^1 dx  \, \frac{2x(1-x)-1}{k^2x(1-x)-m_f^2} \; , 
\\[2ex]
\tilde{S}_{(1)} & \, = \, \sum_{i,f} \,  \frac{\alpha \, m_f^2}{\pi \, v} \, Q_f^2  \, N_C^f  \,\, \text{Im}\big( y^{\varphi_i^0}_f \big) \, \int_0^1 dx  \, \frac{1}{k^2x(1-x)-m_f^2} \; , \\[2ex]
S_{(2)} & \, = \, \sum_i \, \frac{\alpha \, v}{2\pi}   \, \lambda_{\varphi_i^0 H^+H^-} \, \int_0^1 dx  \, \frac{x(x-1)}{k^2x(1-x)-M_{H^\pm}^2} \; ,
\end{align}
for the $\varphi_i^0-\gamma\gamma$ vertices with a fermionic or a charged Higgs loop, in agreement with \cite{Cheung:2009fc}. As for the third diagram, we find
\begin{align}
S_{(3)} & \, = \, \sum_i \,  \frac{\alpha}{2 \pi \, v}  \, \mathcal{R}_{i1} \, \int_0^1 dx  \, \frac{ M_W^2 \, x (3x(4x-1)+10)-M_{\varphi_i^0}^2 \!\ x(1-x)}{k^2x(1-x)-M_W^2} \, .
\end{align}
The new gauge-invariant scalar form factors coming from diagrams (4) to (6) are given by:\vspace{0.3cm} 
\begin{align}
S_{(4)} & \, = \, \frac{\alpha \, N_C \, |V_{tb}|^2 }{2\pi \, v \, s_{\text{w}}} \, \int_0^1 dx  \!\ \frac{ \big[Q_t x + Q_b(1-x)\big] \, \big[\varsigma_u m_t^2 x^2 - \varsigma_d m_b^2 (1-x)^2\big]} {k^2x(1-x)-m_b^2(1-x)-m_t^2 x} \; , \\[2.2ex]
\tilde{S}_{(4)} & \, = \, i \, \frac{\alpha \, N_C \, |V_{tb}|^2}{2\pi v\, s_{\text{w}}} \,  \int_0^1 dx  \, \frac{ \big[Q_t x + Q_b(1-x) \big]\, \big[ -\varsigma_u m_t^2 x + \varsigma_d m_b^2 (x-1)\big]}{k^2x(1-x)-m_b^2(1-x)-m_t^2 x} \; , \\[2.2ex]
S_{(5)} & \, = \, \frac{\alpha}{4\pi \, v \, s_{\text{w}}} \, \sum_i \,   \mathcal{R}_{i1}(\mathcal{R}_{i2} - i\mathcal{R}_{i3})  \!\  \int_0^1 dx \, x^2 \, \,  \frac{(M_{H^\pm}^2 + M_W^2 - M_{\varphi_i^0}^2)(1-x) - 4M_W^2}{k^2 x(1-x) - M_W^2 x- M_{\varphi_i^0}^2(1-x)} \; , \\[2.2ex]
S_{(6)} &= \frac{\alpha \, v}{4\pi s_\mathrm{w}} \, \sum_i \, \lambda_{\varphi_i^0 H^+ H^-} \!\ (\mathcal{R}_{i2} - i \mathcal{R}_{i3} ) \!\  \int_0^1 dx  \!\ \frac{x^2(x-1)}{k^2x(1-x)-M_{H^\pm}^2 x-M_{\varphi_i^0}^2 (1-x)} \; ,
\end{align}
\noindent with $s_\mathrm{w}\equiv\sin\theta_\mathrm{w}$, and $\theta_\mathrm{w}$ the weak mixing angle. 

\subsection{Contributions to $\Delta a_\mu$}

Using the effective vertices from the previous section for calculating the second loop, ignoring suppressed terms proportional to higher powers of $m_\mu^2/M^2$ (with $M$ a $heavy$ mass) in the numerator and the muon mass in the denominator, we obtain the various contributions to the anomalous magnetic moment of the muon. The first two contributions are the well known classical results \cite{Broggio:2014mna,Wang:2014sda,Dedes:2001nx,Gunion:2008dg,Chang:2000ii,Cheung:2001hz,Krawczyk:2002df,Larios:2001ma,Cheung:2003pw,
Arhrib:2001xx,Heinemeyer:2003dq,Kong:2004um,Cheung:2009fc}
\begin{align}
\Delta a_\mu^{(1)}&  =  \sum_{i,f}  \frac{\alpha  \!\ m_\mu^2}{4  \!\  \pi^3  \!\  v^2}  \!\ N_C^f  \!\ Q_f^2  \!\  \Bigg[  \text{Re}\big( y^{\varphi_i^0}_f \big)  \, \,   \text{Re}\big( y^{\varphi_i^0}_l \big) \, \,  \mathcal{F}^{(1)}\bigg(\frac{m_f^2}{M_{\varphi_i^0}^2}\bigg) \, + \, \text{Im}\big( y^{\varphi_i^0}_f  \big)  \, \,   \text{Im} \big( y^{\varphi_i^0}_l \big) \,\,  \tilde{\mathcal{F}}^{(1)}\bigg(\frac{m_f^2}{M_{\varphi_i^0}^2} \bigg)  \Bigg] \, , \\[2.1ex]
\Delta a_\mu^{(2)}& = \sum_i \frac{\alpha  \!\ m_\mu^2}{8  \!\  \pi^3  \!\  M_{\varphi_i^0}^2}  \,\,  \text{Re}\big( y^{\varphi_i^0}_l \big)  \!\ \lambda_{\varphi_i^0 H^+ H^-}  \!\ \mathcal{F}^{(2)}\bigg(\frac{M_{H^\pm}^2}{M_{\varphi_i^0}^2}\bigg) \, .
\end{align}
The third contribution simply reads
\begin{align}
\Delta a_\mu^{(3)}& = \sum_i \frac{\alpha  \!\ m_\mu^2}{8  \!\  \pi^3  \!\  v^2}  \,\,  \text{Re}\big( y^{\varphi_i^0}_l \big)  \!\ \mathcal{R}_{i1}  \!\ \mathcal{F}^{(3)}\bigg(\frac{M_W^2}{M_{\varphi_i^0}^2}\bigg) \, .
\end{align}
As for the new contributions, given by the last three sets in Fig.~\ref{BarZee2}, their contributions are given by
\begin{align}
\Delta a_\mu^{(4)} & = \frac{\alpha  \!\ m_\mu^2  \, N_C \, |V_{tb}|^2}{32  \!\  \pi^3  \!\  s_{\text{w}}^2 \!\ v^2 \, (M_{H^\pm}^2-M_W^2)}   \, \int_0^1 dx  \!\ \Big[ \, Q_t x + Q_b (1-x) \, \Big]  \notag \\ &  \, \, \, \times  \!\   \Big[  \,  \text{Re}(\varsigma_d \varsigma_l^*) \,  m_b^2 x(1-x) + \text{Re}(\varsigma_u\varsigma_l^*)  \, m_t^2x(1+x) \,  \Big] \Bigg[   
\mathcal{G} \bigg( \frac{m_t^2}{M_{H^\pm}^2},\frac{m_b^2}{M_{H^\pm}^2}  \bigg)
-\mathcal{G} \bigg( \frac{m_t^2}{M_{W}^2},\frac{m_b^2}{M_{W}^2}  \bigg)   \Bigg] \, , \label{fermionicContrib} 
\\[2ex]
\Delta a_\mu^{(5)} & =  \frac{\alpha  \!\ m_\mu^2 }{64  \!\  \pi^3  \!\  s_{\text{w}}^2 \!\ v^2 \, (M_{H^\pm}^2-M_W^2)} \, \sum_i  \, \text{Re}\Big[ \varsigma_l^* \!\ \mathcal{R}_{i1}(\mathcal{R}_{i2}-i\mathcal{R}_{i3}) \Big]  \!\  \int_0^1 dx  \!\ x^2  \!\  \notag \\
& \qquad\qquad \times  \!\  \Big[ \!\  \big(M_{H^\pm}^2 + M_W^2 - M_{\varphi_i^0}^2 \big)(1-x)- 4 M_W^2   \!\  \Big] \Bigg[   
\mathcal{G} \bigg( \frac{M_W^2}{M_{H^\pm}^2},\frac{M_{\varphi_i^0}^2}{M_{H^\pm}^2}  \bigg)- \mathcal{G} \bigg( 1,\frac{M_{\varphi_i^0}^2}{M_{W}^2}  \bigg)   \Bigg] \, ,  \\[2.1ex]
\Delta a_\mu^{(6)}&= \frac{\alpha  \, m_\mu^2 }{64  \,  \pi^3  \,  s_{\text{w}}^2 \, (M_{H^\pm}^2-M_W^2) } \, \sum_i  \,   \text{Re}\Big[ \varsigma_l^* \,  (\mathcal{R}_{i2}-i\mathcal{R}_{i3}) \Big] \, \lambda_{\varphi_i^0H^+H^-} \,   \int_0^1 dx  \!\ x^2(x-1)  \notag \\
& \qquad\qquad\qquad\qquad\qquad\qquad\qquad\qquad\qquad\qquad\qquad \times  \!\  \Bigg[   
\mathcal{G} \bigg( 1,\frac{M_{\varphi_i^0}^2}{M_{H^\pm}^2}  \bigg) - \mathcal{G} \bigg( \frac{M_{H^\pm}^2}{M_W^2},\frac{M_{\varphi_i^0}^2}{M_{W}^2}  \bigg) 
   \Bigg] \, .
\end{align}
We can also consider the contribution from a lepton and a neutrino loop by replacing $Q_t\to 0$, $m_t\to 0$, $Q_b \to -1$, $m_b\to m_l$, $\varsigma_d\to \varsigma_l$ and $\varsigma_u \to 0$ in (\ref{fermionicContrib}) and where $m_l$ is the mass of the considered lepton. However, these contributions turn out to be very suppressed due to the smallness of the lepton masses and we shall ignore them in our present analysis. The needed loop functions are given by:

\begin{align}
\mathcal{F}^{(1)}(\omega)&= \frac{\omega}{2}   \int_0^1 dx  \!\  \frac{2x(1-x)-1}{\omega-x(1-x)} \!\ \ln\Bigg(\frac{\omega}{x(1-x)}\Bigg) \, , \\[2ex]
\tilde{\mathcal{F}}^{(1)}(\omega)&= \frac{\omega}{2}   \int_0^1 dx  \!\  \frac{1}{\omega-x(1-x)} \!\ \ln\Bigg(\frac{\omega}{x(1-x)}\Bigg) \, , 
\\[2ex]
\mathcal{F}^{(2)}(\omega)&= \frac{1}{2}   \int_0^1 dx  \!\  \frac{x(x-1)}{\omega-x(1-x)} \!\ \ln\Bigg(\frac{\omega}{x(1-x)}\Bigg) \, , \\[2ex]
\mathcal{F}^{(3)}(\omega)&= \frac{1}{2}   \int_0^1 dx  \!\  \frac{ \!\ x \!\ [3x(4x-1)+10] \omega  - x(1-x)}{\omega-x(1-x)} \!\ \ln\Bigg(\frac{\omega}{x(1-x)}\Bigg) \, ,
\end{align}
and
\begin{align}
\mathcal{G}(\omega^a,\omega^b)&= \frac{\ln\Bigg(\dfrac{\omega^a x + \omega^b(1-x)}{x(1-x)}\Bigg)}{x(1-x)-\omega^ax-\omega^b(1-x)} \, . 
\end{align}

\section{Phenomenology}
\label{sec:phenomeno}

In the present analysis we neglect possible CP-violating effects; {\it i.e.}, we consider a CP-conserving scalar potential and real alignment parameters $\varsigma_f$.
The fermionic couplings of the neutral scalar fields are then given, in units of the SM Higgs couplings, by
\begin{align}  \label{equations1}
&& y_{f}^h & = \cos{\tilde\alpha} + \varsigma_f \sin{\tilde \alpha} \!\ , &&& y_{d,l}^A & =  i\,\varsigma_{d,l}  \!\ , &&  \notag \\
& & y_{f}^H & = -\sin{\tilde\alpha} + \varsigma_f \cos{\tilde \alpha} \!\ , &&&
y_{u}^A \; & =\; -i\, \varsigma_u  \!\  \,,
\end{align}
and the couplings to a pair of gauge bosons (\ref{sumrule}) are simply ($\kappa_V^{\varphi^0_i}\equiv
g_{\varphi^0_iVV}/g_{hVV}^{\mathrm{SM}}$, $V=W,Z$)
\be\label{equations2}
\kappa_V^{h}\;=\; \mathcal{R}_{11} \; = \; \cos{\tilde \alpha} \, , \qquad\qquad
\kappa_V^{H}\;=\; \mathcal{R}_{21} \; = \; -\sin{\tilde \alpha} \, , \qquad\qquad
\kappa_V^{A}\;=\; \mathcal{R}_{31} \; = \;0  \,.
\ee
We shall separate the phenomenological analysis in two parts. For the first part we will analyse the individual contributions from the various $\Delta a_\mu^{(i)}$ factors for different coupling and mass configurations. As for the second part we shall sum all these contributions choosing a few relevant scenarios compatible with collider and flavour bounds and also with constrains from the oblique parameters. Also, we will identify the lightest CP-even Higgs with $h$ and take $M_h=125$ GeV for the whole analysis.

\subsection{Individual $\bf{\Delta a_\mu^{(i)}}$ contributions}

\begin{figure}[!htb]
\centering
\includegraphics[scale=0.6]{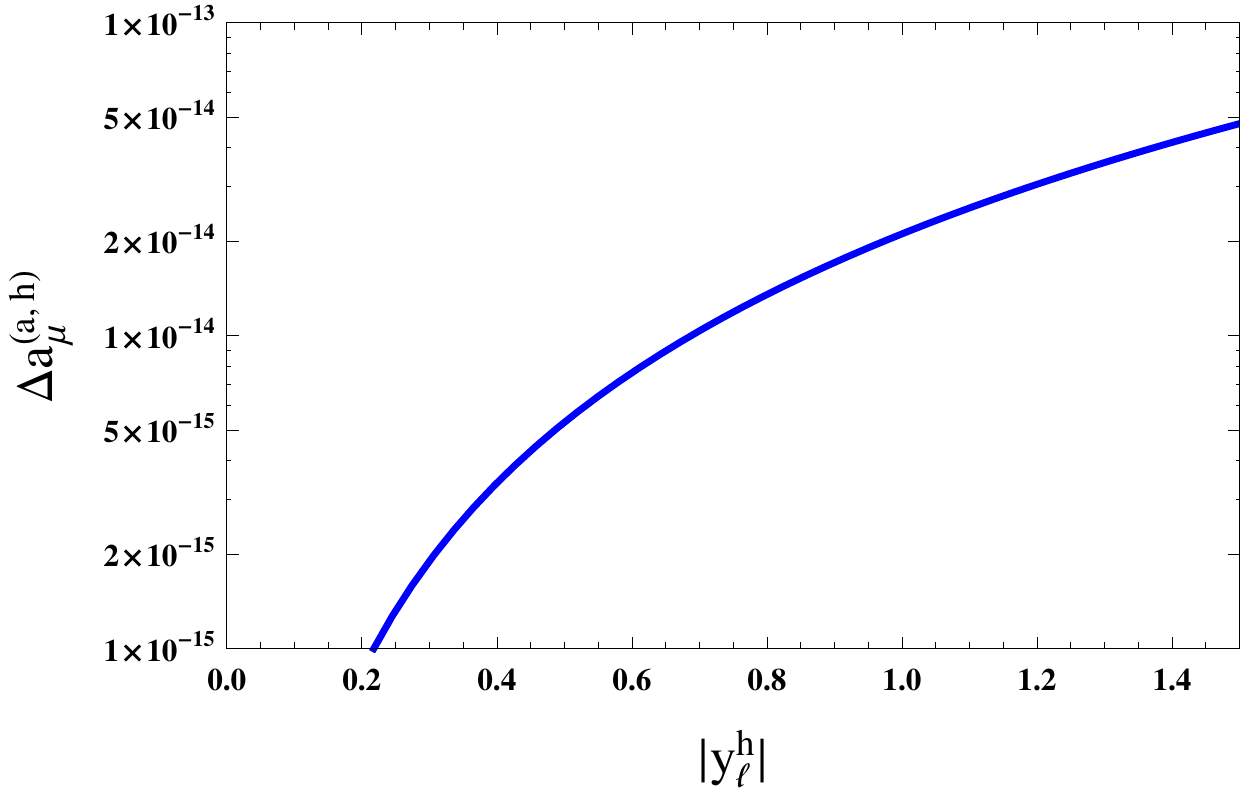} \; \; \includegraphics[scale=0.6]{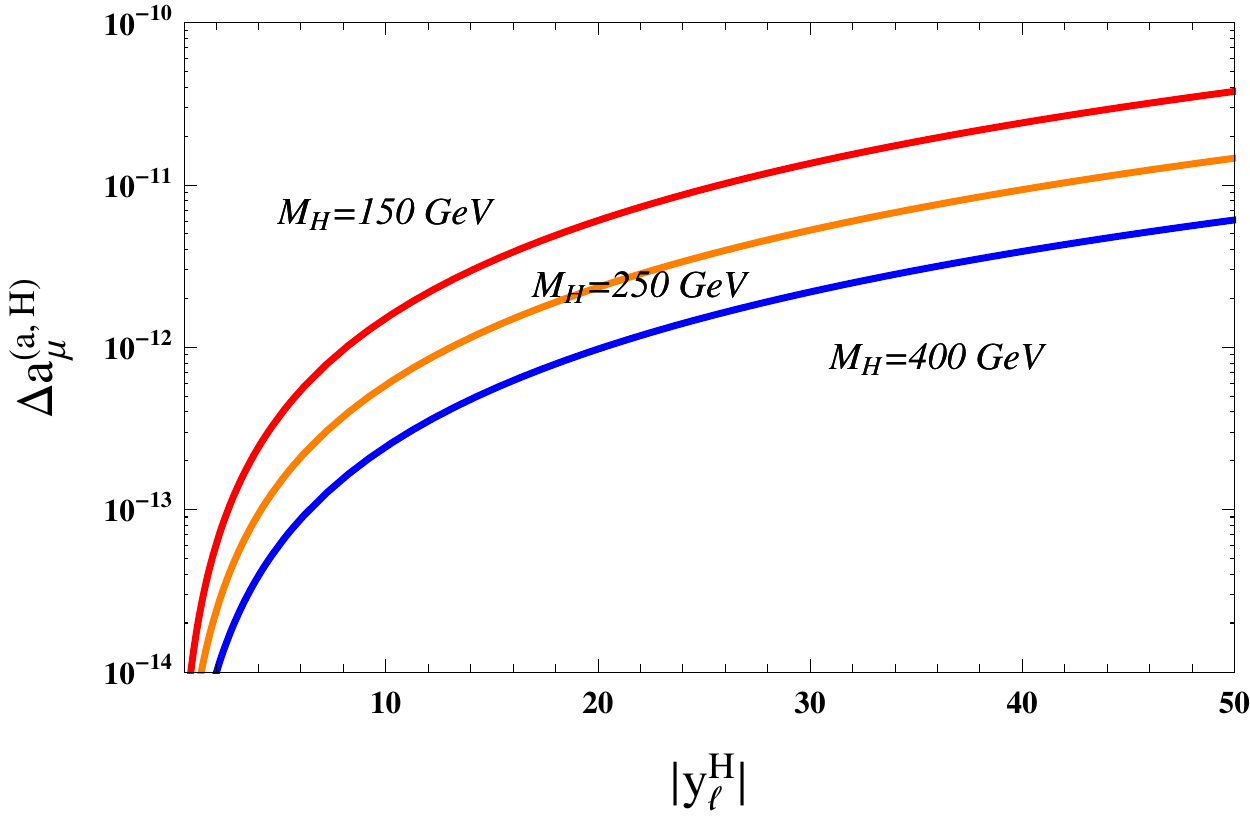} \\[2ex]
\includegraphics[scale=0.6]{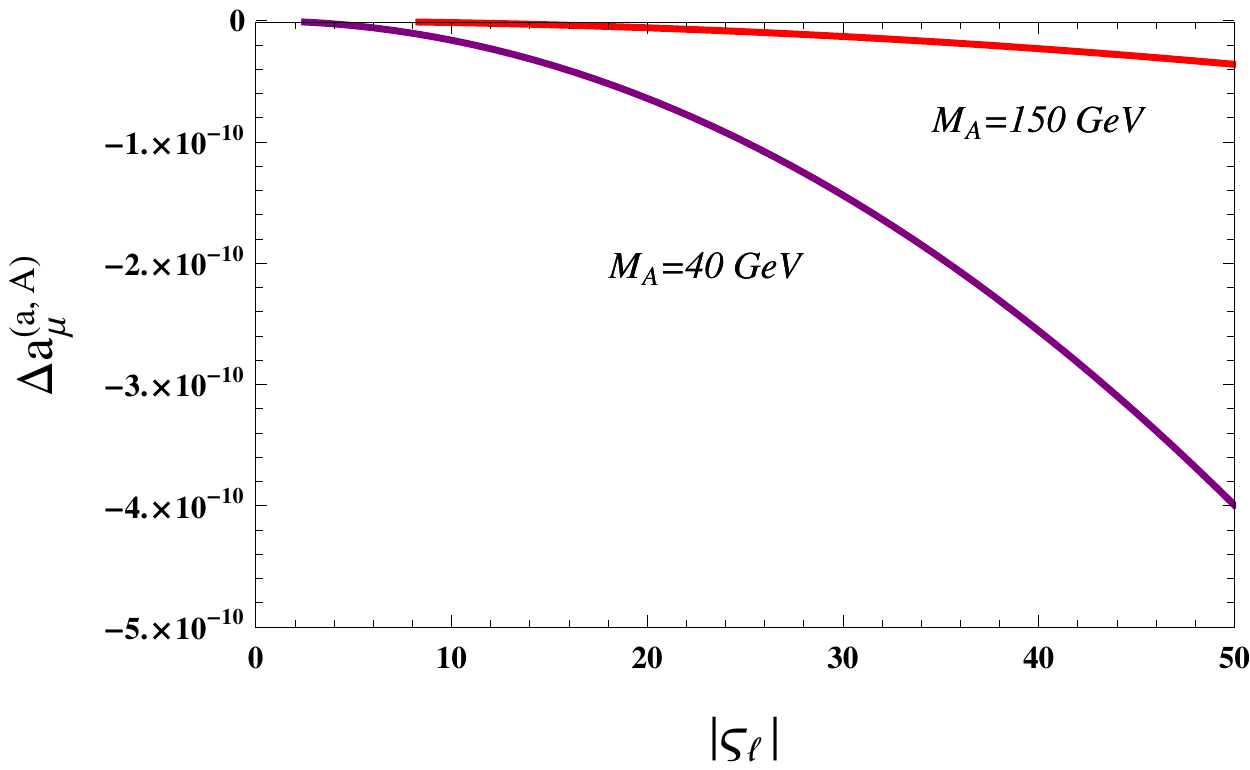} \; \; \includegraphics[scale=0.6]{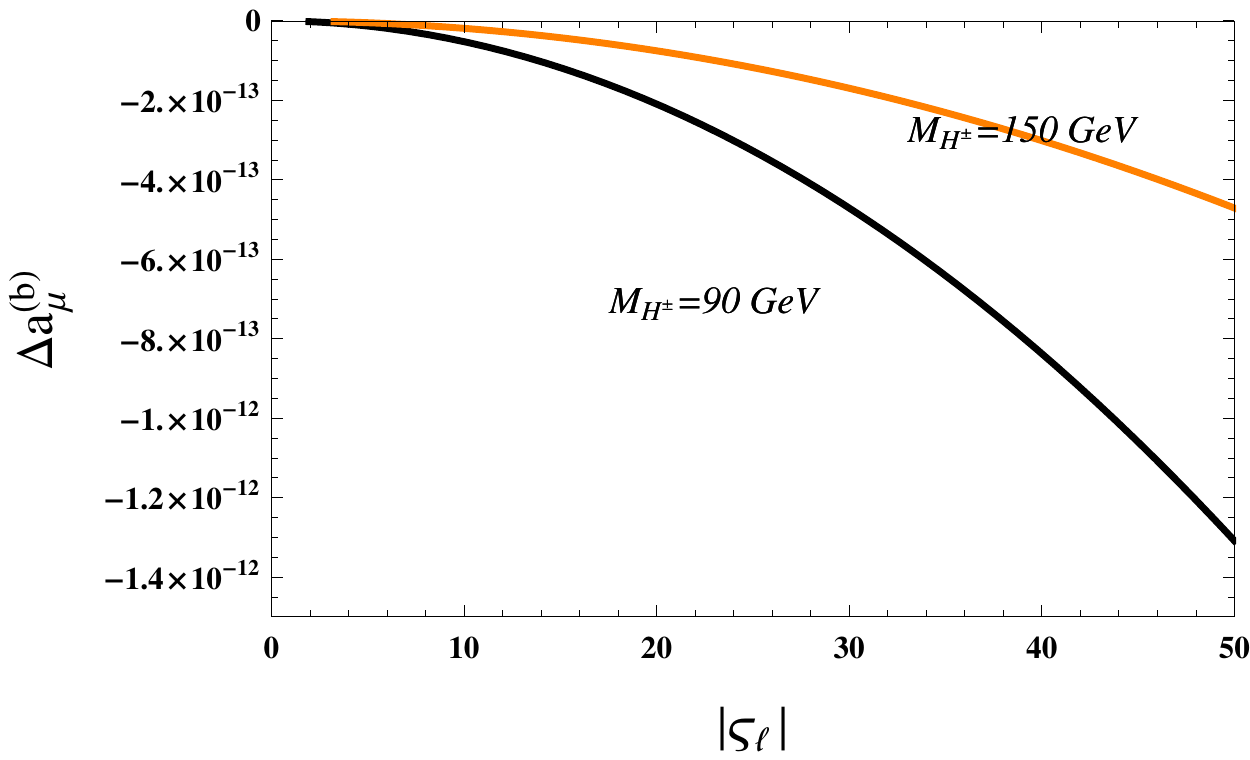}
\caption{{\it One-loop scalar contributions to $\Delta a_\mu$ as functions their couplings to fermions from $h$ (top-left), $H$ (top-right), $A$ (bottom-left) and $H^\pm$ (bottom-right). }}
\label{oneloop}
\end{figure}

\begin{figure}[!htb]
\centering
\includegraphics[scale=0.6]{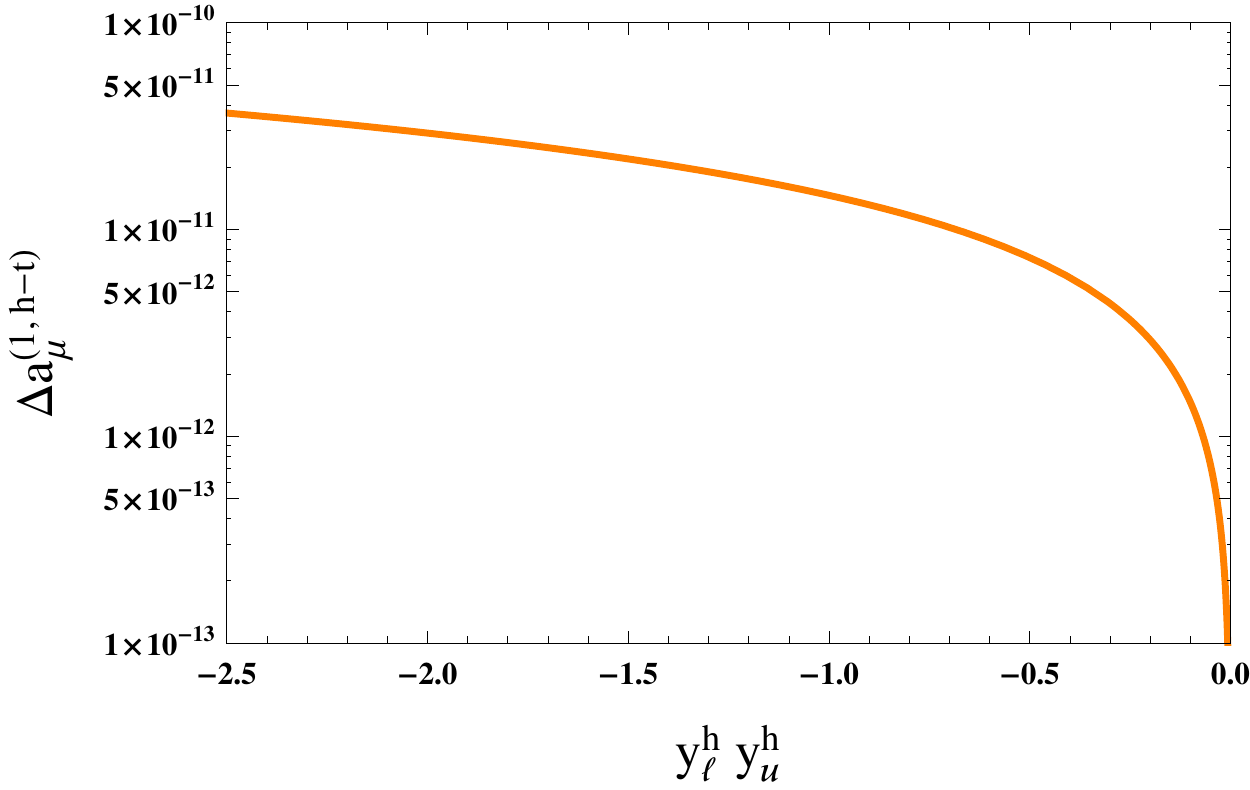} \; \; \includegraphics[scale=0.6]{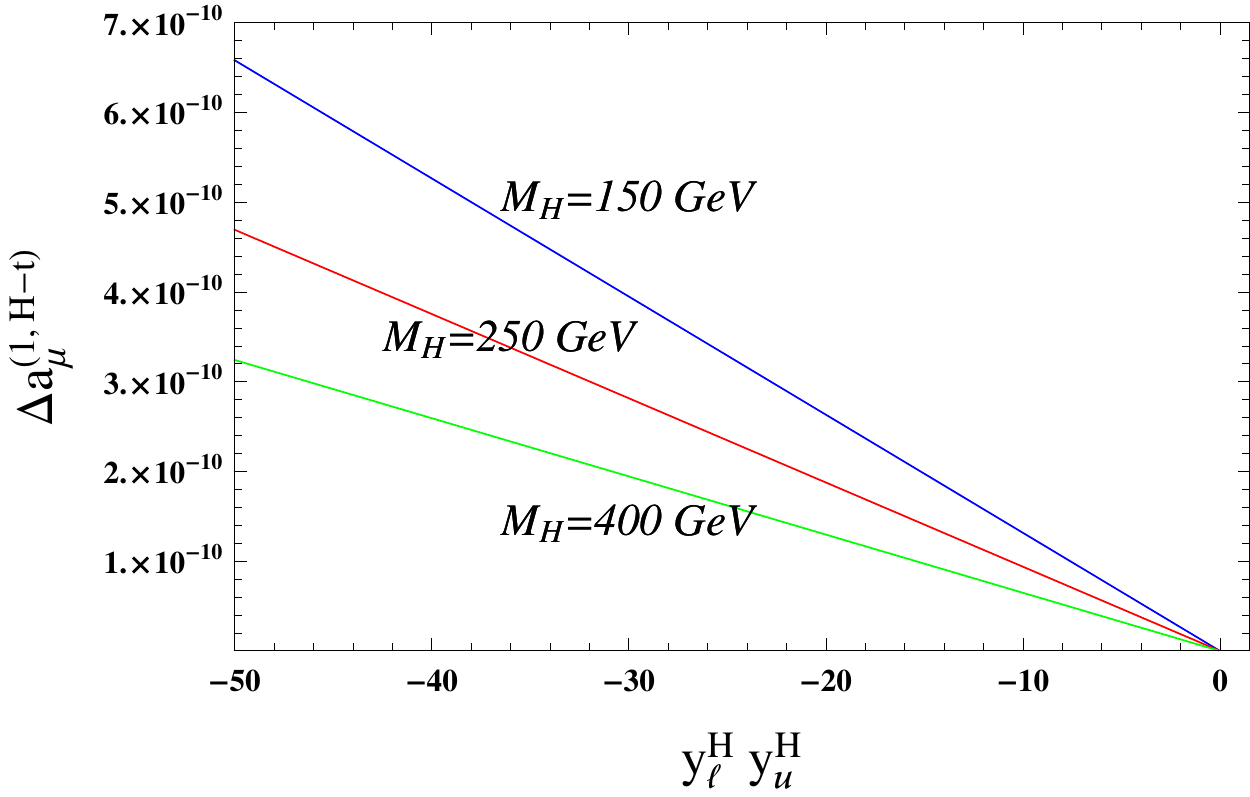} \\[2ex]
\includegraphics[scale=0.6]{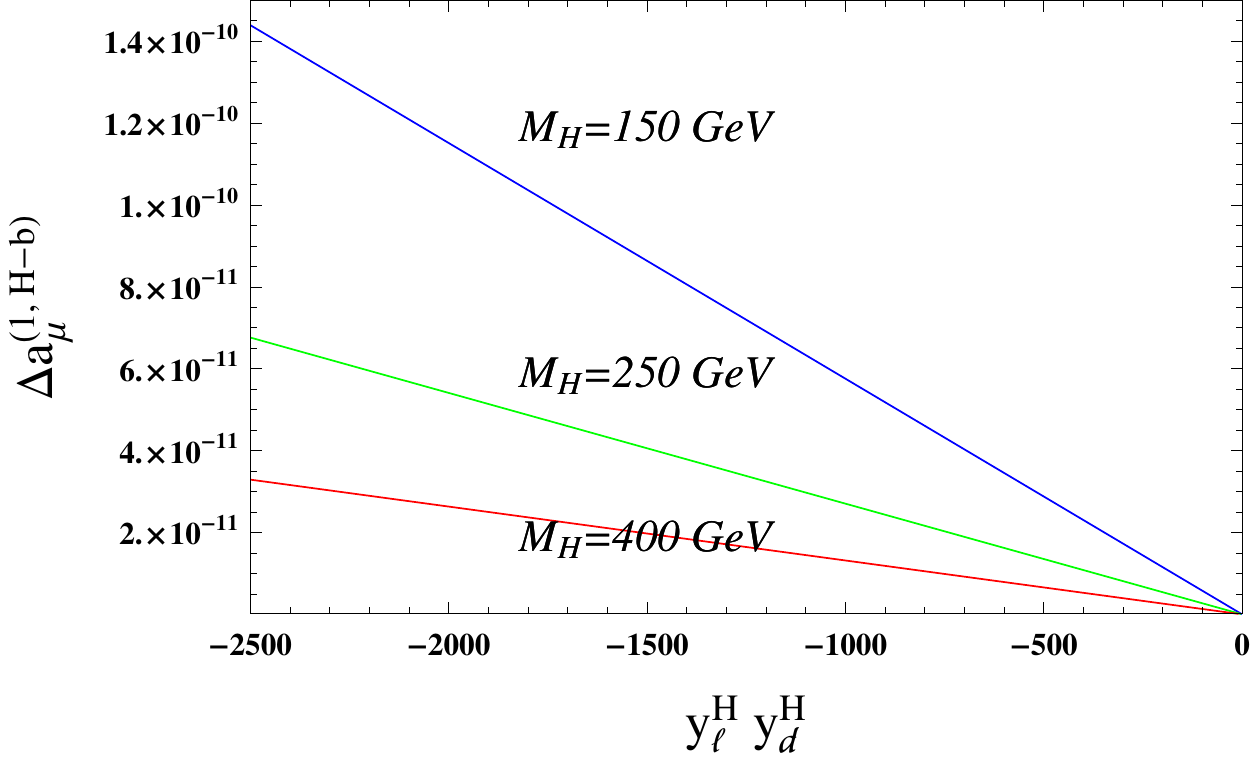} \; \; \includegraphics[scale=0.6]{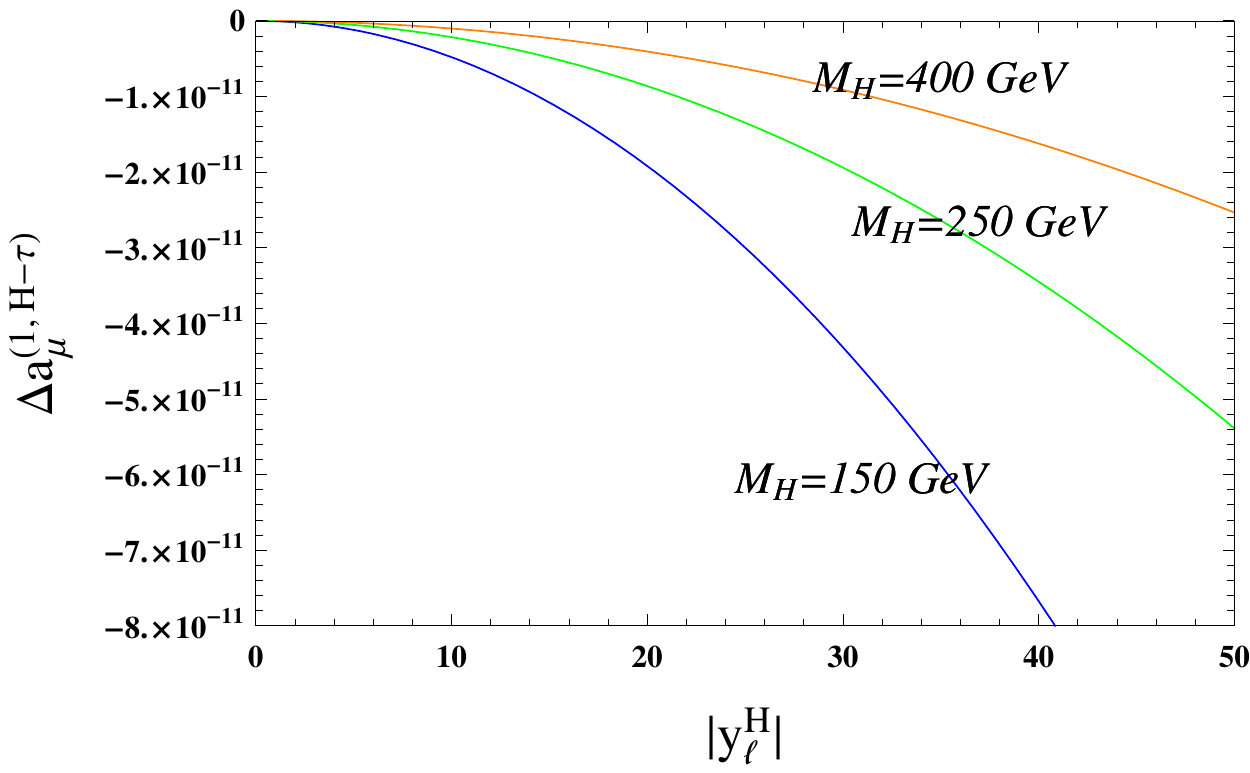}
\caption{{\it Contributions to $\Delta a_\mu^{(1)}$ from $h$ (top-left) and $H$ (top-right) with an associated top-quark loop, and from $H$ with an associated bottom-quark (bottom-left) and tau-lepton (bottom-right) 
loop, as functions of their couplings to fermions.}}
\label{twoloop1hH}
\end{figure}

As we know from global fits to the LHC data, the Yukawa couplings of the discovered scalar boson are SM-like, however with quite large experimental errors. 
The coupling of $h$ to two gauge bosons is constrained by $|\cos\tilde\alpha| > 0.8$ at 95\% CL \cite{ilisie3}. Here we shall always take the positive solution, $\cos\tilde\alpha>0$ (flipping the sign of $\cos\tilde{\alpha}$ leads to an equivalent solution with a sign flip of the couplings $\varsigma_f$). Choosing the positive solution for $\cos\tilde\alpha$, the top Yukawa coupling must also be positive. We shall vary it in the range $y_u^h \in [0.8,1.2]$.
As we know, at least for now, there is no experimental sensitivity to the relative sign of the down-type or leptonic Yukawas with respect to the up-type Yukawas. Therefore we shall be less restrictive with the $y_{d,l}^h$ couplings and allow them to vary in the range $y^h_{d,l} \in [-1.5,1.5]$. As for the alignment parameters, we will vary them as follows: $-1<\varsigma_u<1$ compatible with all flavour constraints and direct charged Higgs searches \cite{ilisie3} for a broad range of the charged Higgs mass, and $-50<\varsigma_{d,l}<50$ to safely avoid the non-perturbative regime.  We shall also vary $y_f^H$ in the same regions as the $\varsigma_f$ parameters (in the limit $\cos\tilde{\alpha}\to 1$ we obtain $y_f^H=\varsigma_f$). The remaining parameters are the couplings of the neutral scalars to a pair of charged Higgses. In order to safely satisfy the perturbativity bounds \cite{ilisie2} for a broad range of $M_{H^{\pm}}$, we will impose $|\lambda_{\varphi_i^0 H^+ H^-}|<5$.

The one-loop well known contribution from the various scalars are shown in Fig.~\ref{oneloop}. The contribution of $h$ is small and positive for the whole considered range of the coupling $|y_l^h|$. The contribution of $H$ is also positive and, its contribution can be of some significance only for large values of $|y_l^H|$ and small values of $M_H$ simultaneously. The contribution of the CP-odd scalar is negative and it is only relevant for large values of $|\varsigma_l|$ and low values of it mass mass, similar to the previous case. As for the charged Higgs contribution, it is always negative and very small, thus irrelevant, at the one loop level.

The two-loop results are presented next. The contribution of $h$, associated with a top-quark loop, to $\Delta a_\mu^{(1)}$ is shown in Fig.~\ref{twoloop1hH} (top-left). It is positive for $y_l^h \, y_u^h <  0$. The contribution of the same scalar $h$ associated with bottom and tau loops is much smaller, of $\mathcal{O}(10^{-13})$ or less for the whole considered parameter space, and is not shown here. The contribution of $H$ for different mass configurations and for different fermionic loops is also shown in 
Fig.~\ref{twoloop1hH}. This contribution is proportional to the $y_l^H$ coupling which can be large. Thus is turns out to be non-negligible even for the sub-dominant bottom-quark and tau-lepton loops. The top-quark loop contribution can be large for all considered mass settings as long as $y_l^H$ is large, and it is positive for $y_l^H \, y_u^H <  0$, as we can observe in 
Fig.~\ref{twoloop1hH} (top-right). The bottom-quark loop contribution can be additionally enhanced by the coupling $y_d^H$, thus, it can overcome the mass suppression. This contribution is positive for $y_l^H \, y_d^H <  0$, see Fig.~\ref{twoloop1hH} (bottom-left). Similar considerations about the enhancement factor $(y_l^H)^2$ can be made for the tau-lepton part, however this contribution is always negative, as shown in the bottom-right panel of Fig.~\ref{twoloop1hH}.

\begin{figure}[!htb]
\centering
\includegraphics[scale=0.6]{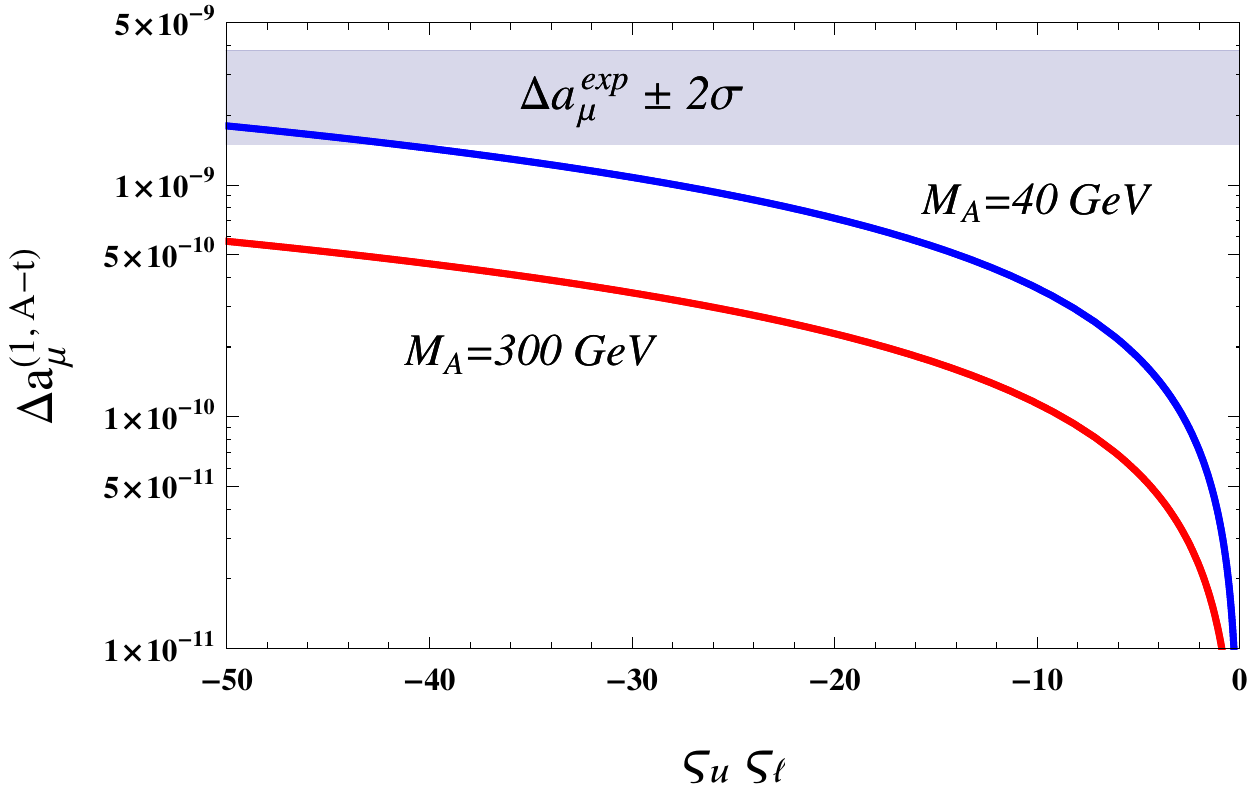} \; \; \includegraphics[scale=0.6]{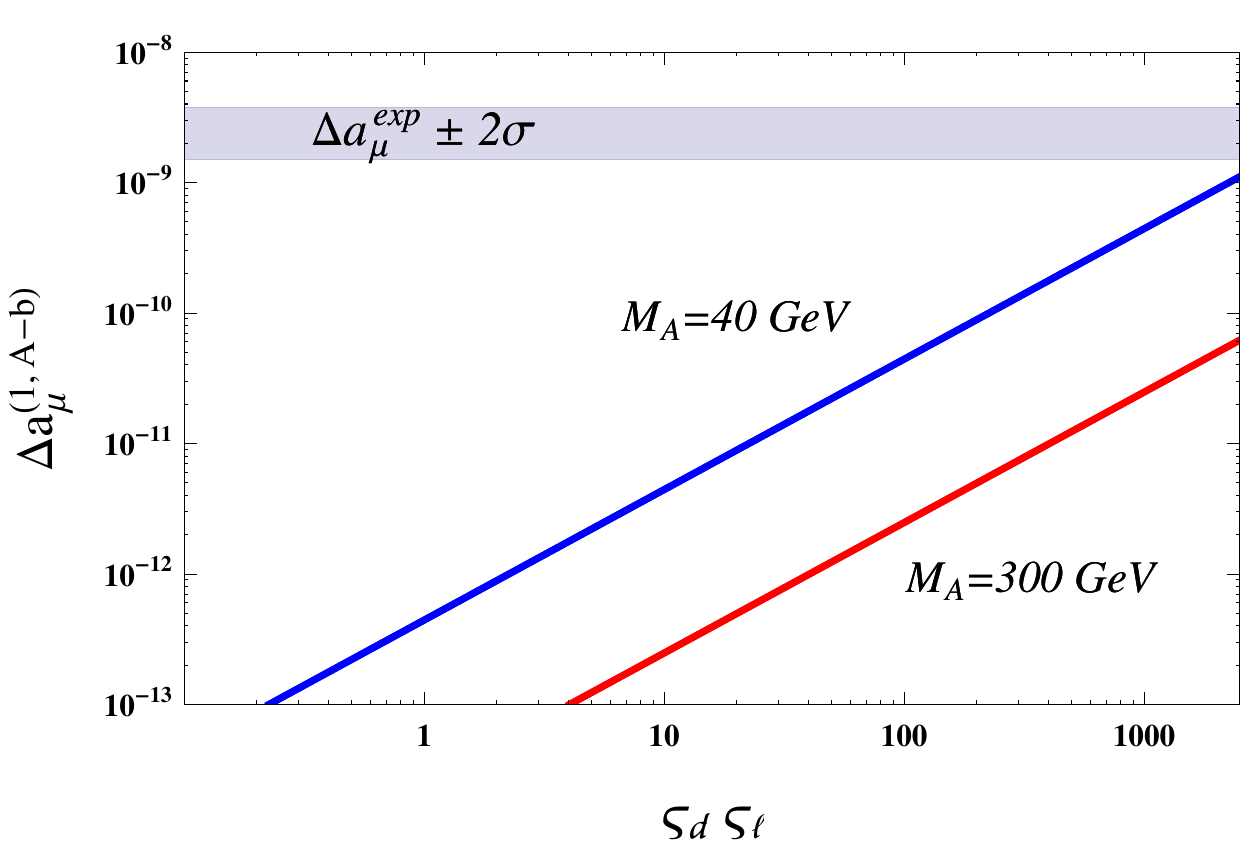}
\caption{{\it Contributions to $\Delta a_\mu^{(1)}$ from the CP-odd scalar $A$, associated with a top-quark (left) and bottom-quark (right) loop, as functions its couplings to fermions. }}
\label{twoloop1A}
\end{figure}

\begin{figure}[!htb]
\centering
\includegraphics[scale=0.6]{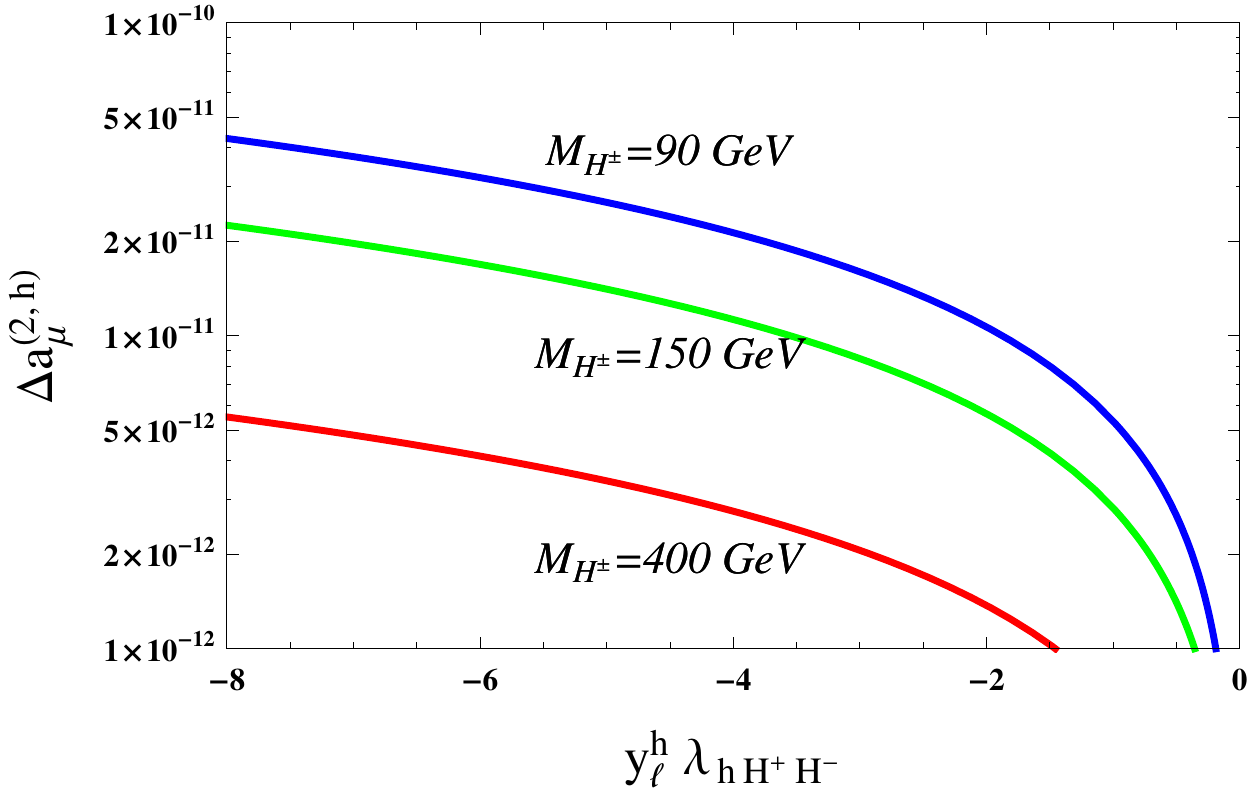} \; \; \includegraphics[scale=0.6]{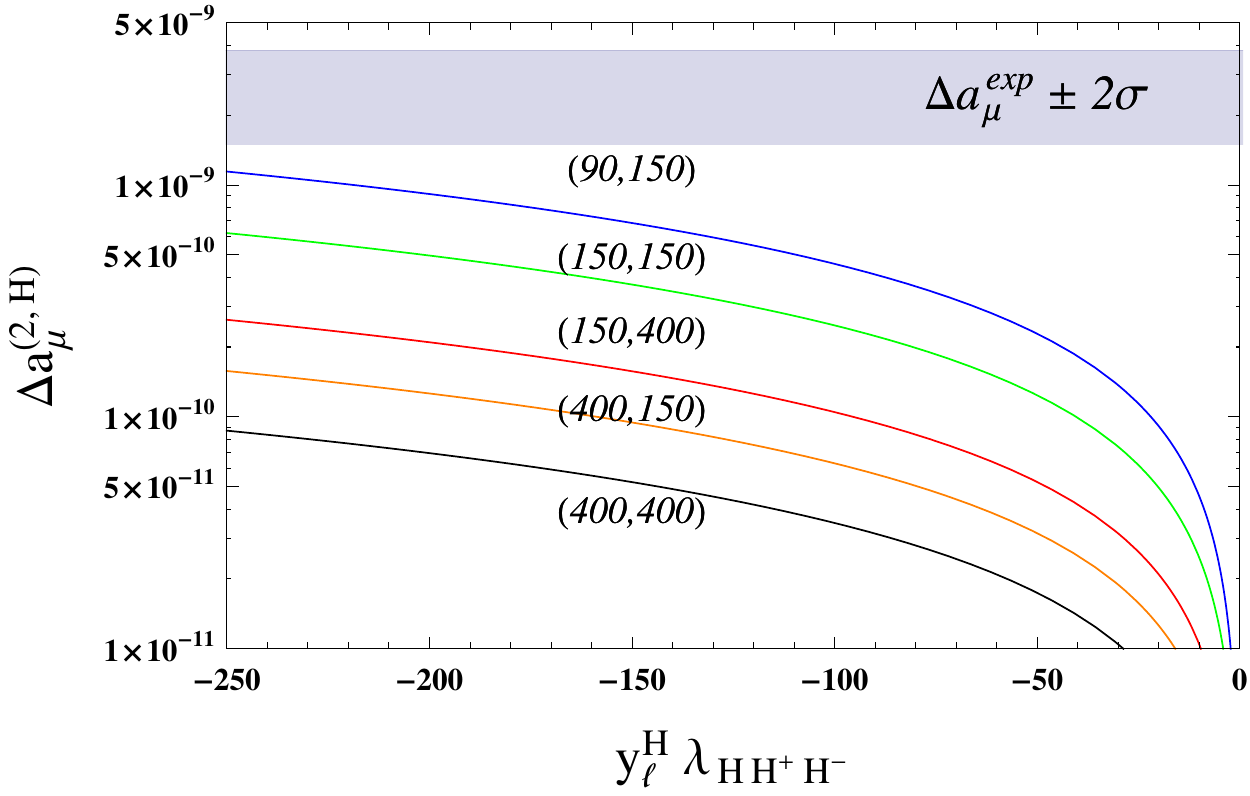}
\caption{{\it Contributions to $\Delta a_\mu^{(2)}$ from $\varphi_i^0={h,H}$ as functions of the product of the couplings
$y_l^{\varphi_i^0} \; \lambda_{\varphi_i^0 H^+ H^-}$ for various charged Higgs masses (left) and for various configurations of 
$(M_{H^\pm},\, M_H)$ (GeV) (right).}}
\label{twoloop2hH}
\end{figure}

The contribution of the CP-odd scalar to $\Delta a_\mu^{(1)}$ is probably the most interesting yet. It has been extensively analysed in previous works \cite{Broggio:2014mna,Wang:2014sda,Dedes:2001nx,Gunion:2008dg,Chang:2000ii,Cheung:2001hz,Krawczyk:2002df,Larios:2001ma,Cheung:2003pw}. For low values of its mass and large values of $\varsigma_{d,l}$ it can reach values within or close to the two-sigma region of 
$\Delta a_\mu^{exp}$, as it is plotted in Fig.~\ref{twoloop1A}. Its value is positive for $\varsigma_u \, \varsigma_l<0 \; (\varsigma_d \, \varsigma_l>0)$ for the top (bottom) quark loop contribution and is always positive for the tau loop contribution. This last case is not shown. It is worth mentioning, however that the tau loop contribution is somewhat larger than the (absolute value of the) bottom contribution. Even if the tau-lepton has a relative mass suppression, the bottom-quark has a charge suppression that is in general larger.

\begin{figure}[!htb]
\centering
\includegraphics[scale=0.6]{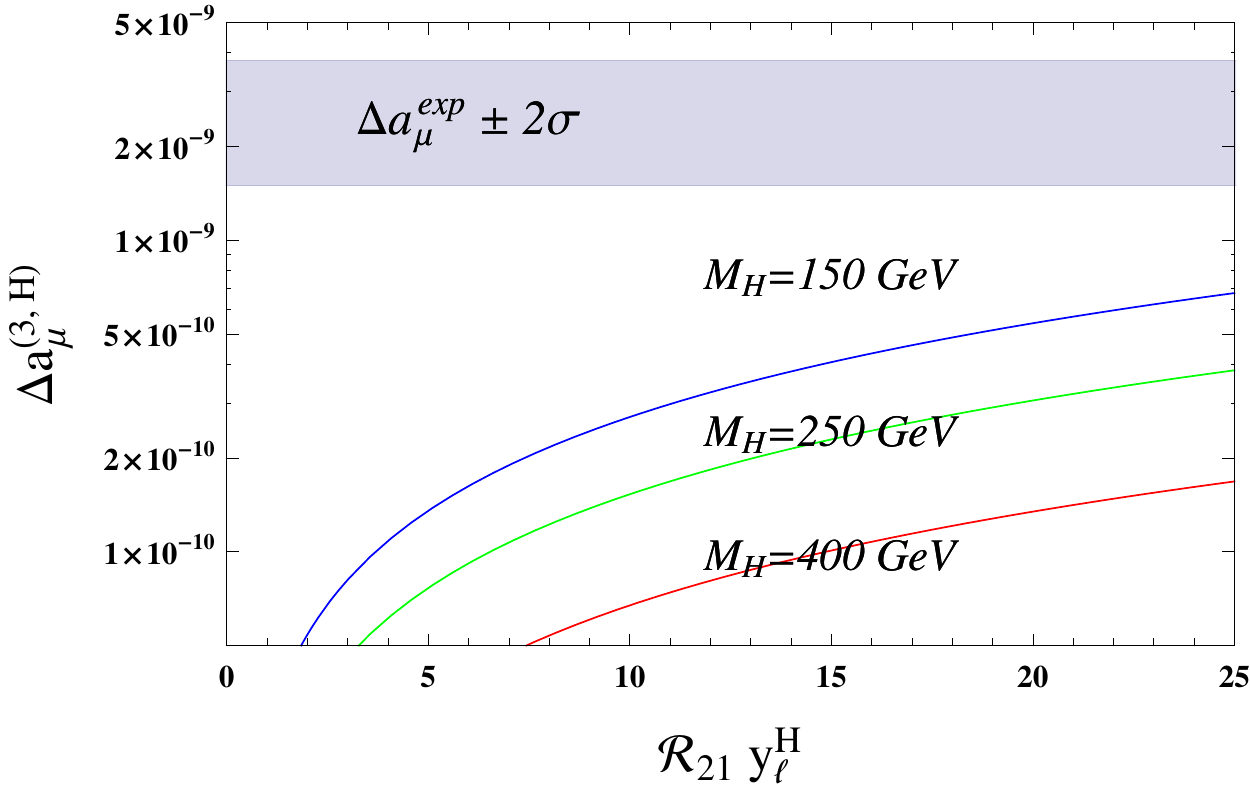} \; \; \includegraphics[scale=0.6]{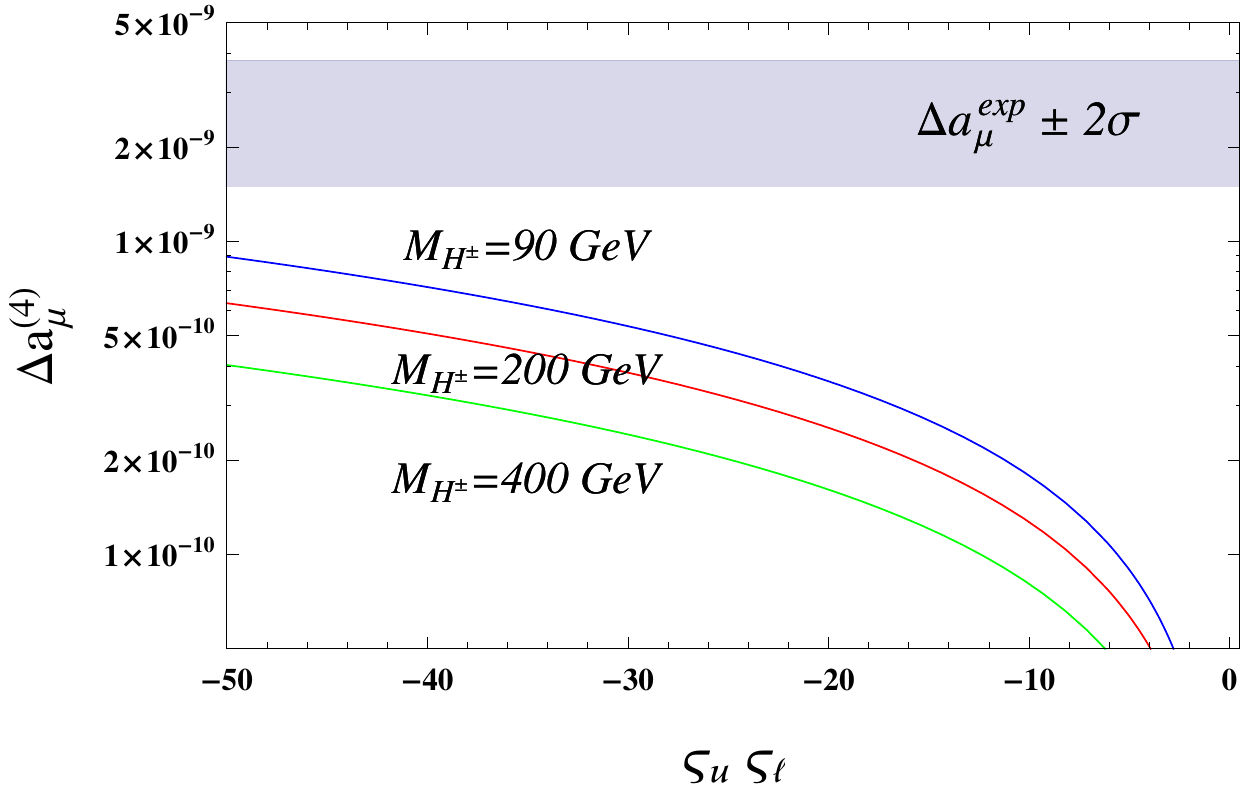}
\caption{{\it Contribution to $\Delta a_\mu^{(3)}$ from $H$ (left) as function of the product of the couplings
$\mathcal{R}_{21} \, y_l^H$ for various mass configurations. Two-loop dominant contribution from the top-bottom quark loops to $\Delta a_\mu^{(4)}$ (right).}}
\label{twoloop3H}
\end{figure}

For $\Delta a_\mu^{(2)}$ we only have two possible contributions, from $h$ and $H$ (in the CP-conserving limit the vertex $AH^+H^-$ vanishes \cite{ilisie2}). The contribution of the light scalar $h$
is relatively small for the whole considered parameter space, Fig.~\ref{twoloop2hH} (left) and that is due to the fact that $y_l^h \in [-1.5,1.5]$ whereas $y_l^H$ can be much larger. The contribution of $H$ can be quite large depending on the configuration of the masses $(M_{H^\pm}, \, M_H)$ (GeV). It reaches its largest value for low masses of both $M_{H^\pm}$ and $M_H$ and large values of the product of the couplings $y_l^H \, \lambda_{HH^+H^-}$. However, even for lower values of the couplings but with low masses (or large masses and large couplings) the contribution can be non-negligible. For details see Fig.~\ref{twoloop2hH} (right).

The next contribution we focus on is $\Delta a_\mu^{(3)}$. The contribution from the light scalar is small, of $\mathcal{O}(10^{-11})$ or less (after subtracting the SM), therefore we can safely neglect it. The $H$ contribution however, is non-negligible. It reaches higher values (and it is positive) for low values of $M_H$ and large positive values of the product $\mathcal{R}_{21} \, y_l^H \, (= \sin^2\tilde{\alpha}-\varsigma_l \sin\tilde{\alpha} \, \cos\tilde{\alpha})$ as it is plotted in Fig.~\ref{twoloop3H} (left). As we have already mentioned before, this diagram should not be neglected, as it can introduce sizeable effects for some regions of the parameter space.

Now we move on to the analysis of the charged Higgs contributions of the Barr-Zee type diagrams (Fig.~\ref{BarZee2}), which is the main goal of this paper.
It is obvious from Fig.~\ref{twoloop3H} (right) that the $\Delta a_\mu^{(4)}$ contribution is non-negligible for a large region of the parameter space, except for very small values of the product $|\varsigma_u \, \varsigma_l|$. For a charged Higgs with a low mass, say 90 GeV, and large negative values of $\varsigma_l \, \varsigma_u$ this contribution alone can explain around 35 $\%$ of the measured discrepancy. This looks very appealing, because with the exception of a very light CP-odd scalar, the previous contributions cannot reach such large values. For the plot shown in Fig.~\ref{twoloop3H} (right) we have chosen $\varsigma_d=0$. However, a variation of $\varsigma_d$ in its allowed interval $[-50,50]$ only produces a shift in the plotted values of order $10^{-12}$ or less. This is obviously due to a relative suppression factor $m_b^2/m_t^2$ and therefore this contribution can be safely ignored.

\begin{figure}[!htb]
\centering
\includegraphics[scale=0.6]{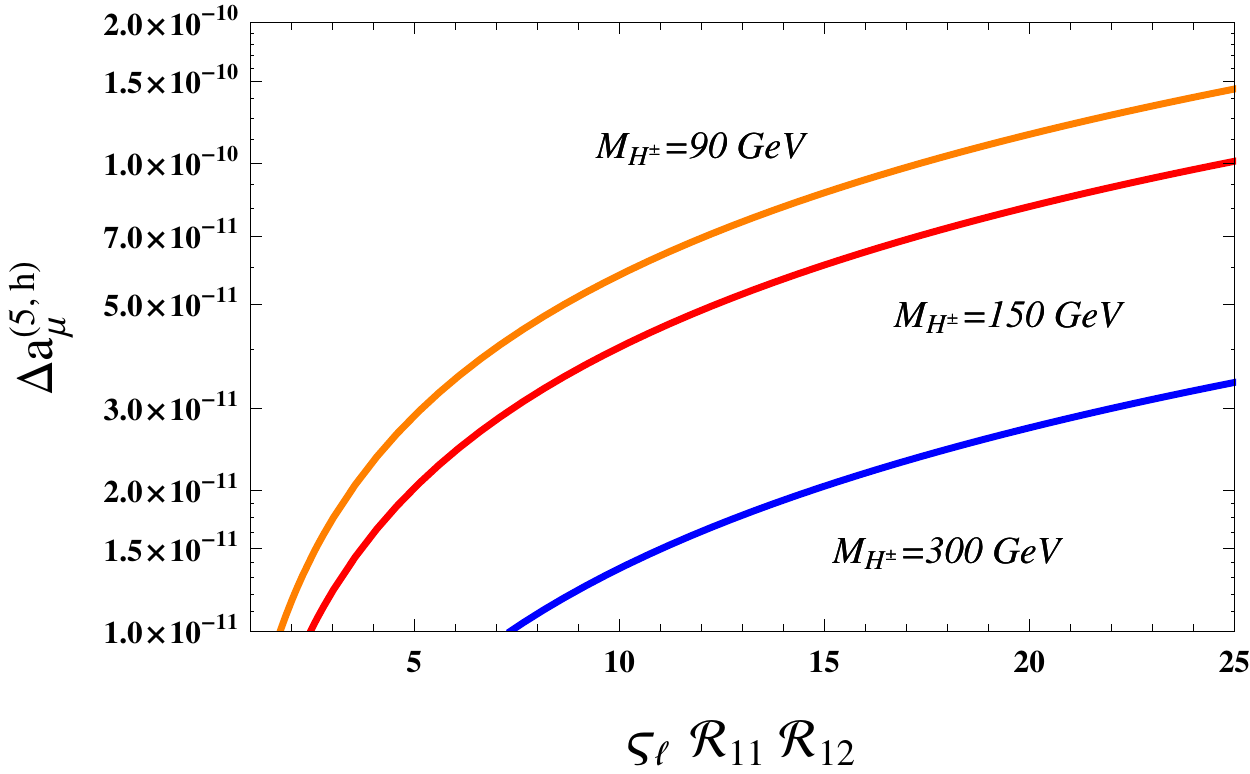} \; \; \includegraphics[scale=0.6]{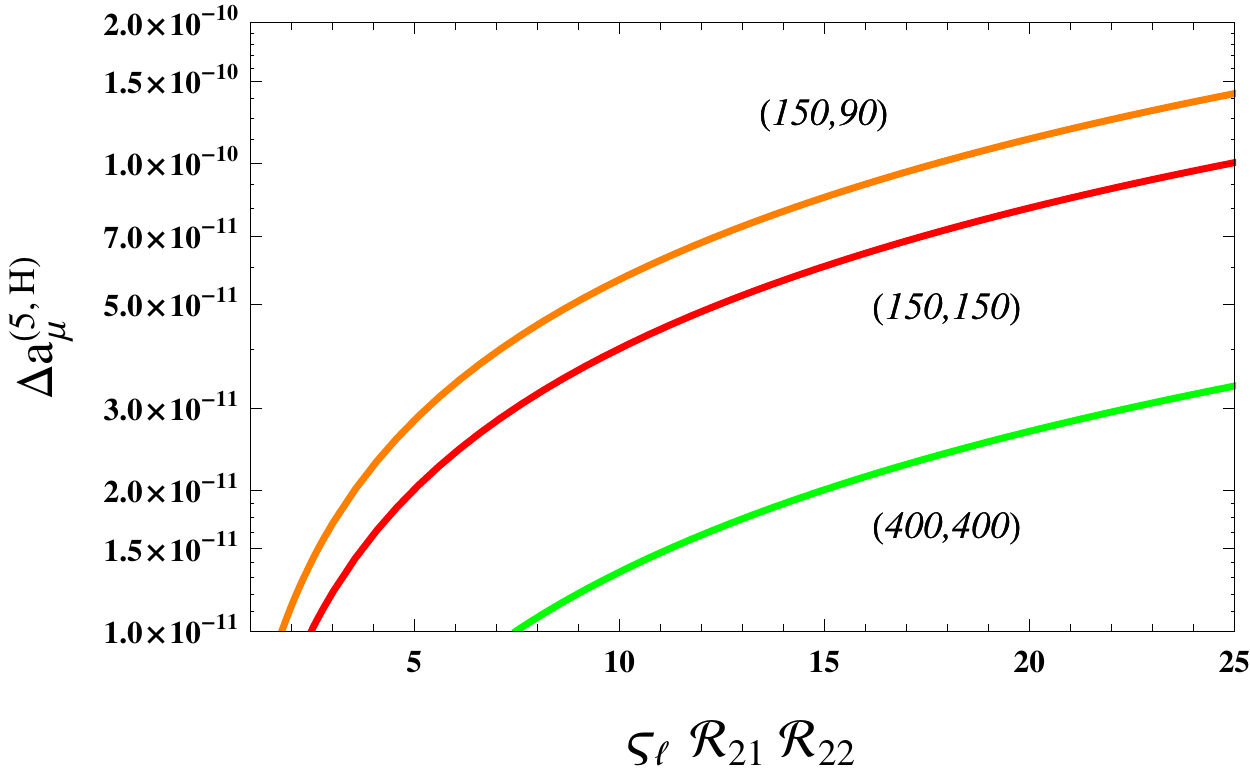}
\caption{{\it Contributions to $\Delta a_\mu^{(5)}$ from $\varphi_i^0={h,H}$ as functions of the product of the couplings
$\varsigma_l \, \mathcal{R}_{i1} \, \mathcal{R}_{i2}$ for various charged Higgs masses (left) and for various configurations of 
$(M_H,\, M_{H^\pm})$ (GeV) (right).}}
\label{twoloop5}
\end{figure}

\begin{figure}[!htb]
\centering
\includegraphics[scale=0.6]{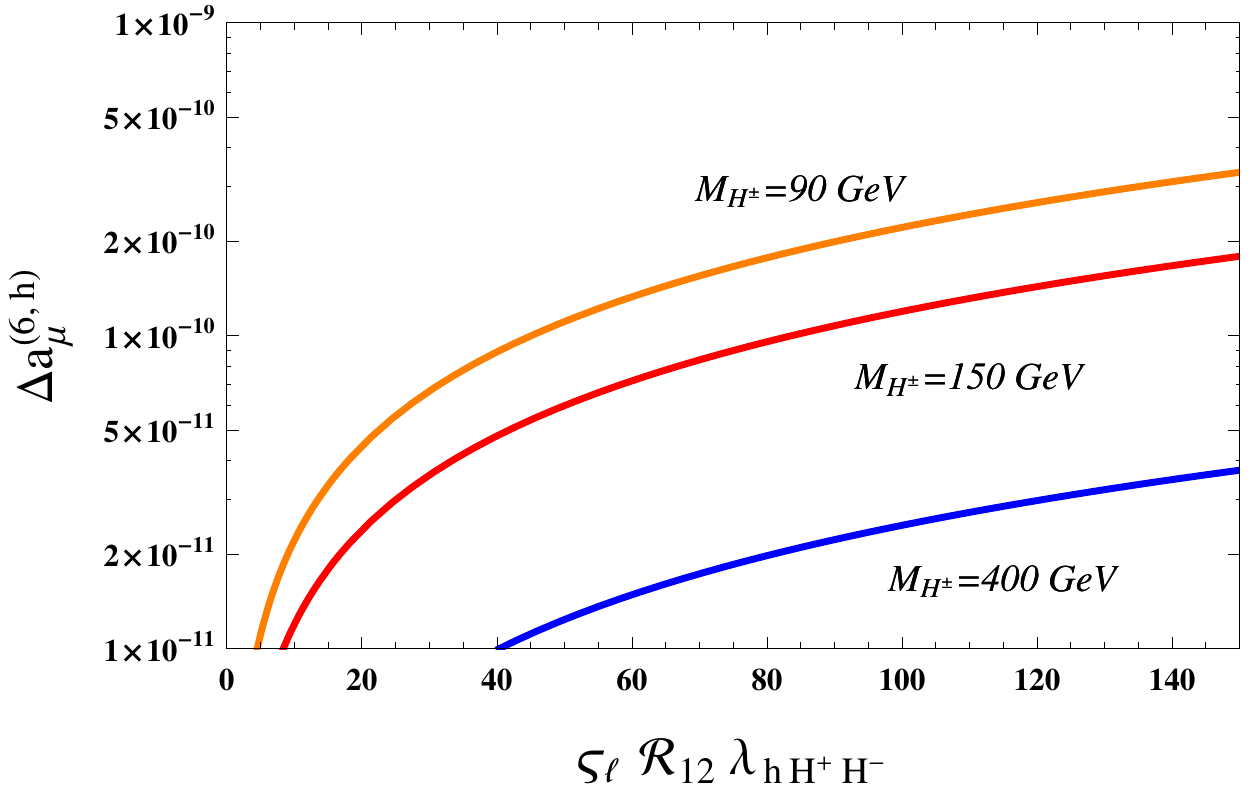} \; \; \includegraphics[scale=0.6]{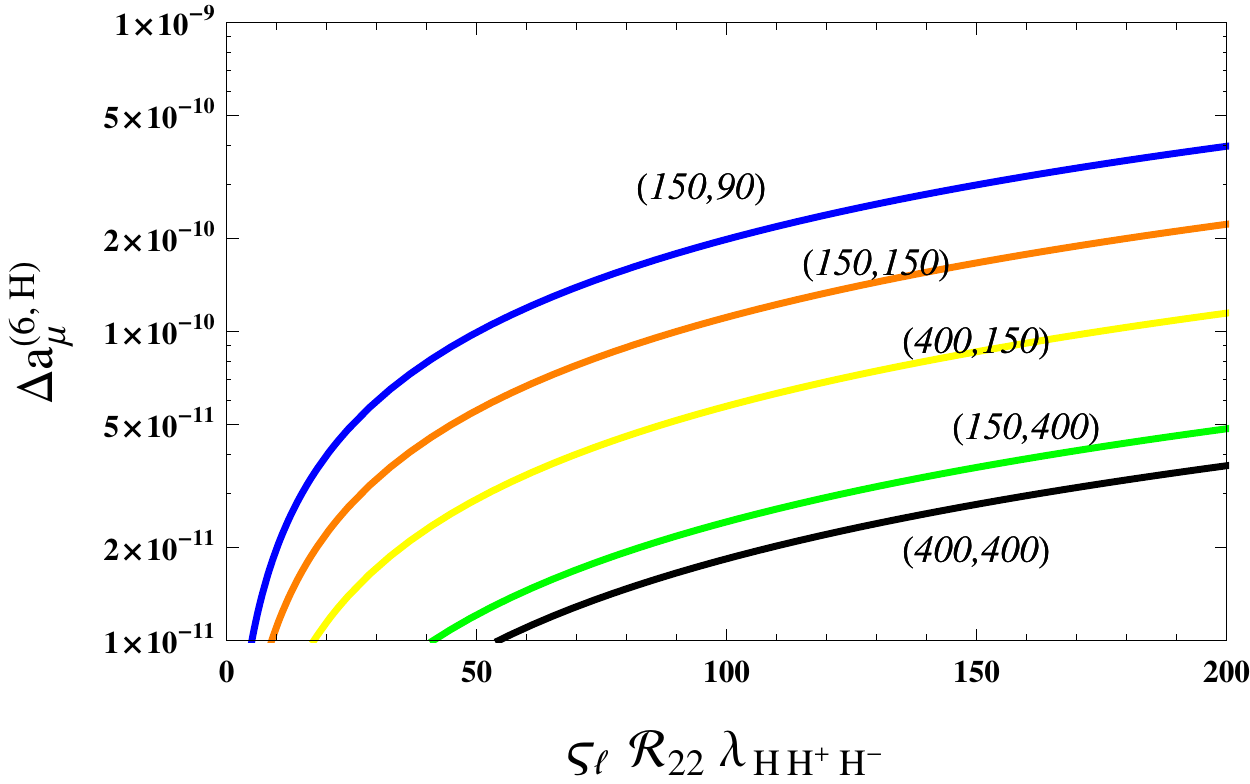}
\caption{{\it Contributions to $\Delta a_\mu^{(6)}$ from $\varphi_i^0={h,H}$ as functions of the product of the couplings
$\varsigma_l \, \mathcal{R}_{i2} \, \lambda_{\varphi_i^0 H^+ H^-}$ for various charged Higgs masses (left) and for various configurations of 
$(M_H,\, M_{H^\pm})$ (GeV) (right).}}
\label{twoloop6}
\end{figure}

Last, contributions $\Delta a_\mu^{(5)}$ and $\Delta a_\mu^{(6)}$ are shown in Fig.~\ref{twoloop5} and Fig.~\ref{twoloop6}. They are a little bit smaller, however they can reach values up to $10^{-10}$. Again this happens, for small mass configurations and large values of the corresponding couplings. We can see in Fig.~\ref{twoloop5} that both $h$ and $H$ contributions 
can be very similar, however, they cannot be simultaneously positive (if the product of the three couplings $\varsigma_l \, \mathcal{R}_{i1} \, \mathcal{R}_{i2}$ is chosen positive for one scalar, for the other must necessarily be negative). On the other hand, both $h$ and $H$ contributions from $\Delta a_\mu^{(6)}$ can be simultaneously positive, and of similar value. Thus, when summed up they can play an important role in the total value of $\Delta a_\mu$.

We have proven thus, that these new Barr-Zee contributions must not be ignored, as they might sizeably modify the theoretical prediction for this observable within the 2HDM framework.

\subsection{Total contribution to $\bf{(g-2)_\mu}$}

\begin{figure}[!htb]
\centering
\includegraphics[scale=0.65]{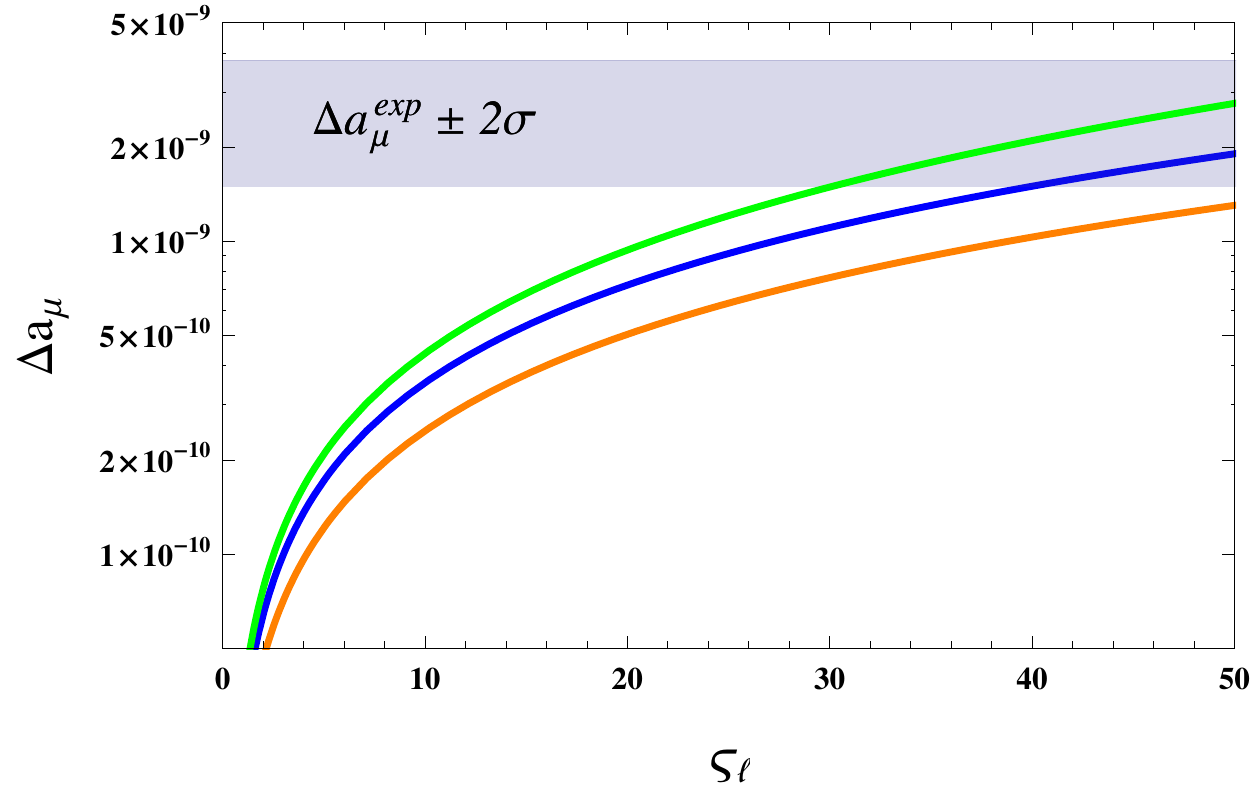} \;\; \includegraphics[scale=0.65]{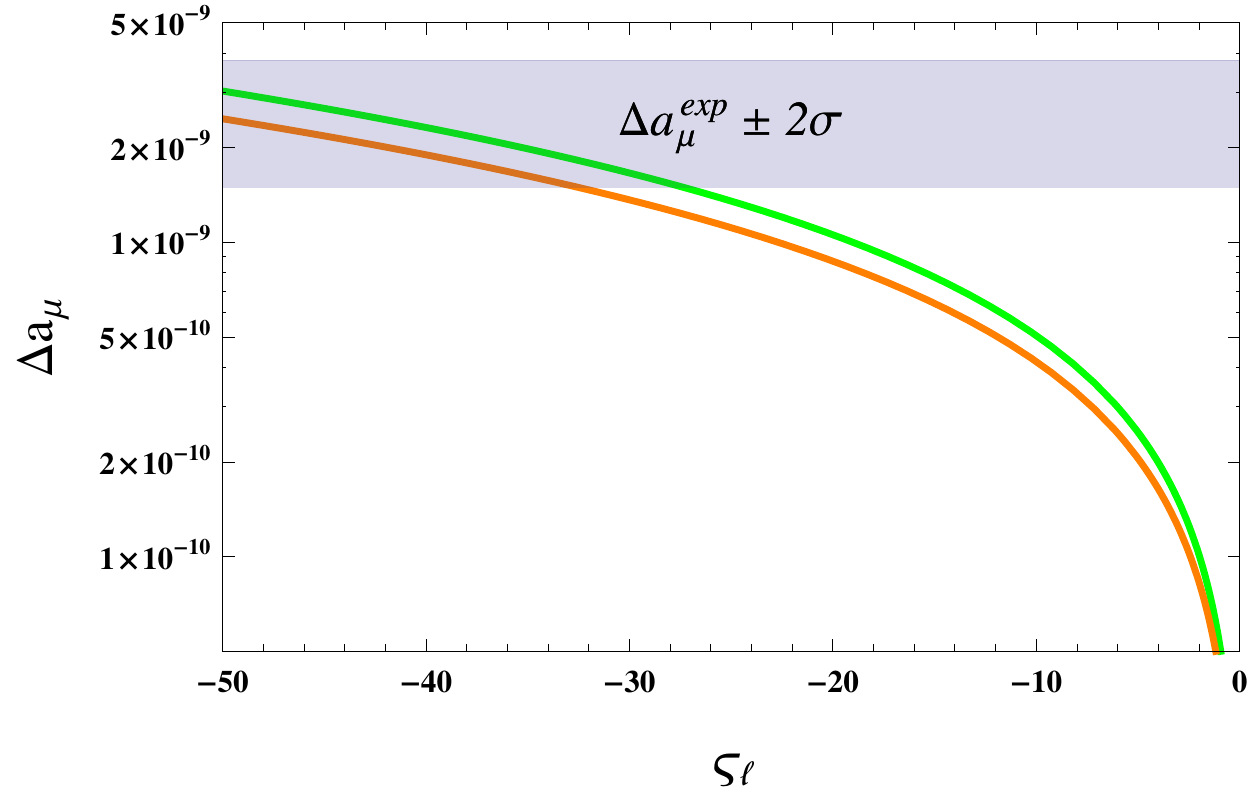}  
\caption{{\it Total $\Delta a_\mu$ contribution as a function of $\varsigma_l$ for different coupling and mass configurations.}}
\label{h_125_ab}
\end{figure}

Thus, we have seen that the dominant contributions of the new Bar-Zee type diagrams come from the mechanisms (3) (Fig.~\ref{BarZee1}) and (4) (Fig.~\ref{BarZee2}). All the other $new$ contributions are sub-dominant. Now, it is interesting to put all these results together, and show the total effect on $\Delta a_\mu$ for a few relevant scenarios.  In Fig.~\ref{h_125_ab} (left panel) we show $\Delta a_\mu$ as a function of $\varsigma_l$ for positive values of this coupling and for a few scenarios given by $\cos\tilde\alpha=0.9$, $\varsigma_u=-0.8$, $\varsigma_d=-20$, 
$M_h=125$ GeV, $\lambda_{h H^+ H^-}=0$, $\lambda_{h H^+ H^-}=-5$. The masses (in GeV) of the remaining scalars are chosen the following way: $M_H=M_{H^{\pm}}=M_A=250$ (lower orange curve), 150 (middle blue curve), $M_H=M_{H^{\pm}}=150$ and $M_A=50$ (upper green curve). Similar to the previous case, but this time for negative values of $\varsigma_l$, in the right panel we have chosen the following parameter configuration: $\cos\tilde\alpha=0.9$, $\varsigma_u=0.8$, $\varsigma_d=2$, 
$M_h=125$ GeV, $\lambda_{h H^+ H^-}=0$, $\lambda_{h H^+ H^-}=5$ and $M_H=M_{H^{\pm}}=250$ GeV and $M_A=40$ GeV (upper green curve) or  $M_H=M_{H^{\pm}}=350$ GeV and $M_A=50$ GeV (lower orange curve).
As expected, from the analysis of the various $\Delta a_\mu^{(i)}$ individual contributions, one obtains a significant contribution for low masses of the scalars (especially for low $M_A$) and large couplings. 
We can also observe that in some cases we do not need the maximum allowed value of $|\varsigma_l|$ in order to reach the two-sigma region of $\Delta a_\mu^{exp}$; a value around $|\varsigma_l| \sim 30$ might just be enough.

\section{Conclusions}
\label{sec:conclusions}

It is a common belief that only a restrained number of diagrams, namely (1) and (2) from Fig.~\ref{BarZee1}, can significantly contribute to $\Delta a_\mu$ in 2HDMs and in most of the previous analyses \cite{Broggio:2014mna,Wang:2014sda,Dedes:2001nx,Gunion:2008dg,Chang:2000ii,Cheung:2001hz,Krawczyk:2002df,Larios:2001ma,Cheung:2003pw}, a CP-odd scalar in the low-mass range is enough to explain, or reduce, the discrepancy between theory and experiment. In this work we have shown that the extra degrees of freedom of the A2HDM given by the $\varsigma_f$ parameters, can also explain this discrepancy in some region of the parameter space, and if not, they can significantly reduce it in most cases. We have also seen that the $W$ loop contribution associated with a heavy scalar $H$ (diagram (3) from Fig.~\ref{BarZee1}) can bring important contributions even if it has a global suppression factor $\mathcal{R}_{21}$. This contribution is positive for negative values of $\varsigma_l$. The most interesting case is, however, the fermionic loop contribution (diagrams (4) from Fig.~\ref{BarZee2}) with the dominant part given by the top-quark.
The last two diagrams (5) and (6) are also interesting, as they can sum up to an $\mathcal{O}(10\%)$ of the total contribution. 
Also, we have seen that not all of these new contributions can be made simultaneously positive, however the total $\Delta a_\mu$ is positive
 for most parameter configurations.  

A highly interesting scenario, that we defer for future work, is to consider CP-violating effects. The imaginary part of the parameters of the potential and especially of the Yukawa sector might be able to bring somewhat sizeable effects.

%%%%%%%%%%%%%%%%%%%%%%%%%%%%%%%%%%%%%%%%%%%%%%%%%%%%%%%%%%%%%%%%%%%%%%%%%%%%%%%%%%%%%%%%%%%%%%%%%%%%%%%%%%%%%%%%%%%%%%%%%%
%%%%%%%%%%%%%%%%%%%%%%%%%%%%%%%%%%%%%%%%%%%%%%%%%%%%%%%%%%%%%%%%%%%%%%%%%%%%%%%%%%%%%%%%%%%%%%%%%%%%%%%%%%%%%%%%%%%%%%%%%%
%%%%%%%%%%%%%%%%%%%%%%%%%%%%%%%%%%%%%%%%%%%%%%%%%%%%%%%%%%%%%%%%%%%%%%%%%%%%%%%%%%%%%%%%%%%%%%%%%%%%%%%%%%%%%%%%%%%%%%%%%%

\begin{appendix}

\section{$WW\gamma$ effective vertex contribution to $(g-2)_\mu$}
\label{appendixa}

\begin{figure}[tbp]
\centering
\includegraphics[scale=0.45]{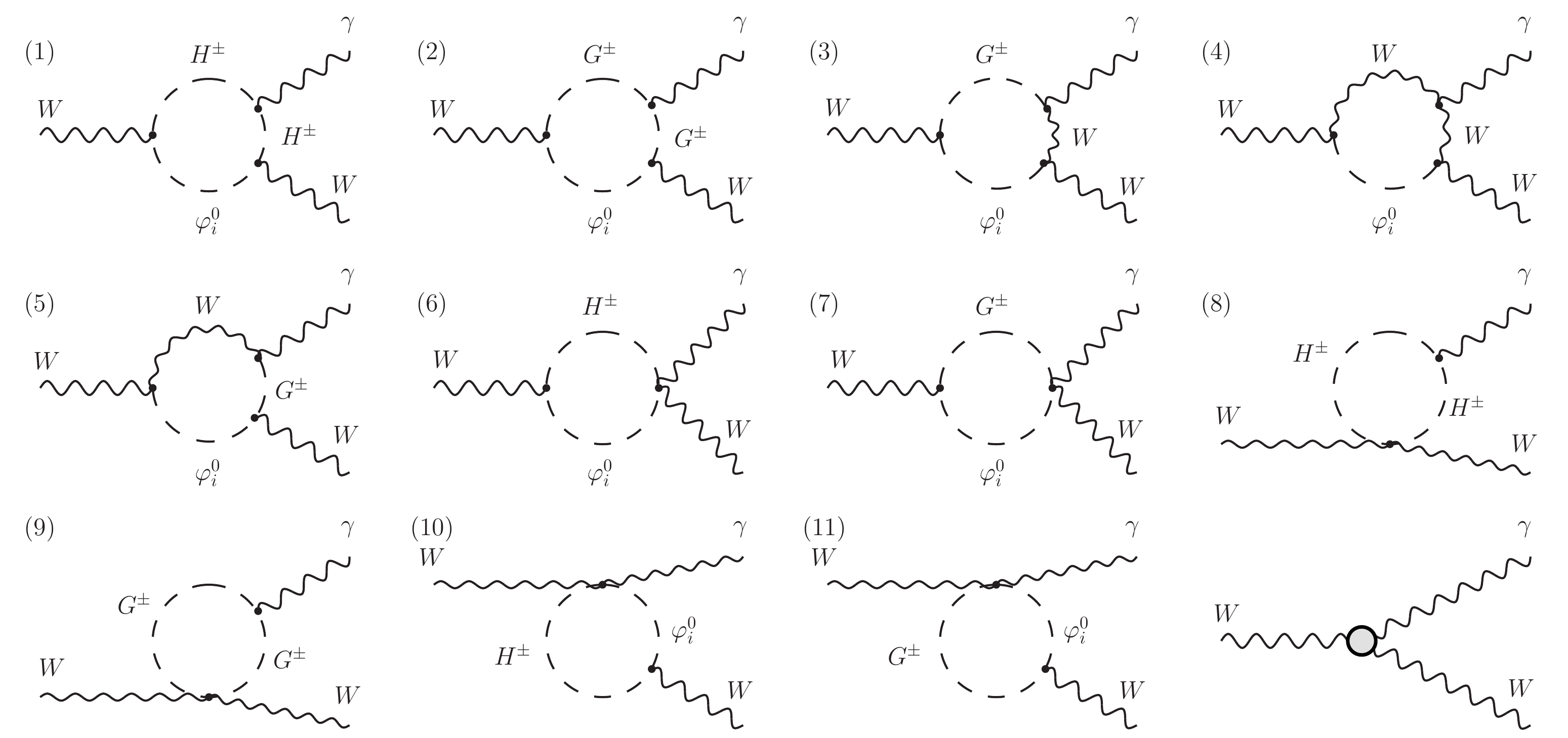}
\caption{{\it One-loop contributions to the $WW\gamma$ effective vertex. The last diagram stands for the one-loop counter-term.}}
\label{WWgeffV}
\end{figure}

\begin{figure}[tbp]
\centering
\includegraphics[scale=0.45]{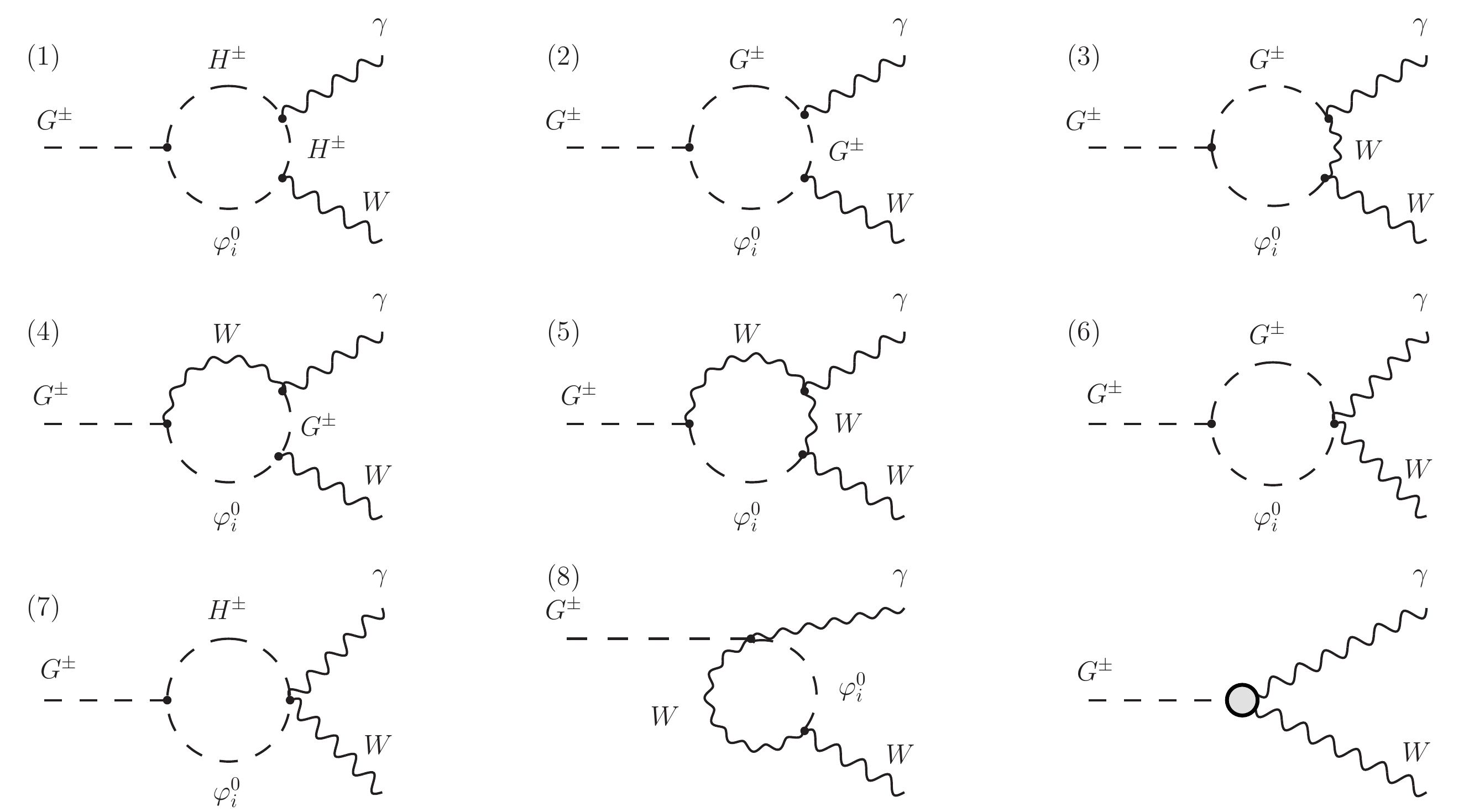}
\caption{{\it One-loop contributions to the $WG^\pm\gamma$ effective vertex. The last diagram stands for the one-loop counter-term.}}
\label{WGgeffV}
\end{figure}

In this section we present the explicit calculation of the contributions from Fig.~\ref{BarZeeAB} (B) to $(g-2)_\mu$.
The 2HDM contributions to the one-loop $WW\gamma$ effective vertex are shown in Fig.~\ref{WWgeffV}, where last diagram stands for the one-loop renormalization counter-term. For this calculation we have followed the renormalization prescription described in \cite{Santos:1996vt}.
Following this prescription one does not need to renormalize the gauge-fixing Lagrangian. Thus, we simply worked in the Feynman gauge \cite{ilisie2}. Working in this gauge, one also needs to take into account $WG^\pm\gamma$ (Fig.~\ref{WGgeffV}) and $G^\pm G^\mp \gamma$ effective vertices. The last set ($G^\pm G^\mp \gamma$) will give rise to contributions to the anomalous magnetic moment that will have a relative suppression factor of $m_\mu^2/M_W^2$ (just as in case (A) of Fig.~\ref{BarZeeAB} for the $H^\pm H^\mp \gamma$ effective vertex), and therefore will not be taken into account.

\begin{figure}[tbp]
\centering
\includegraphics[scale=0.475]{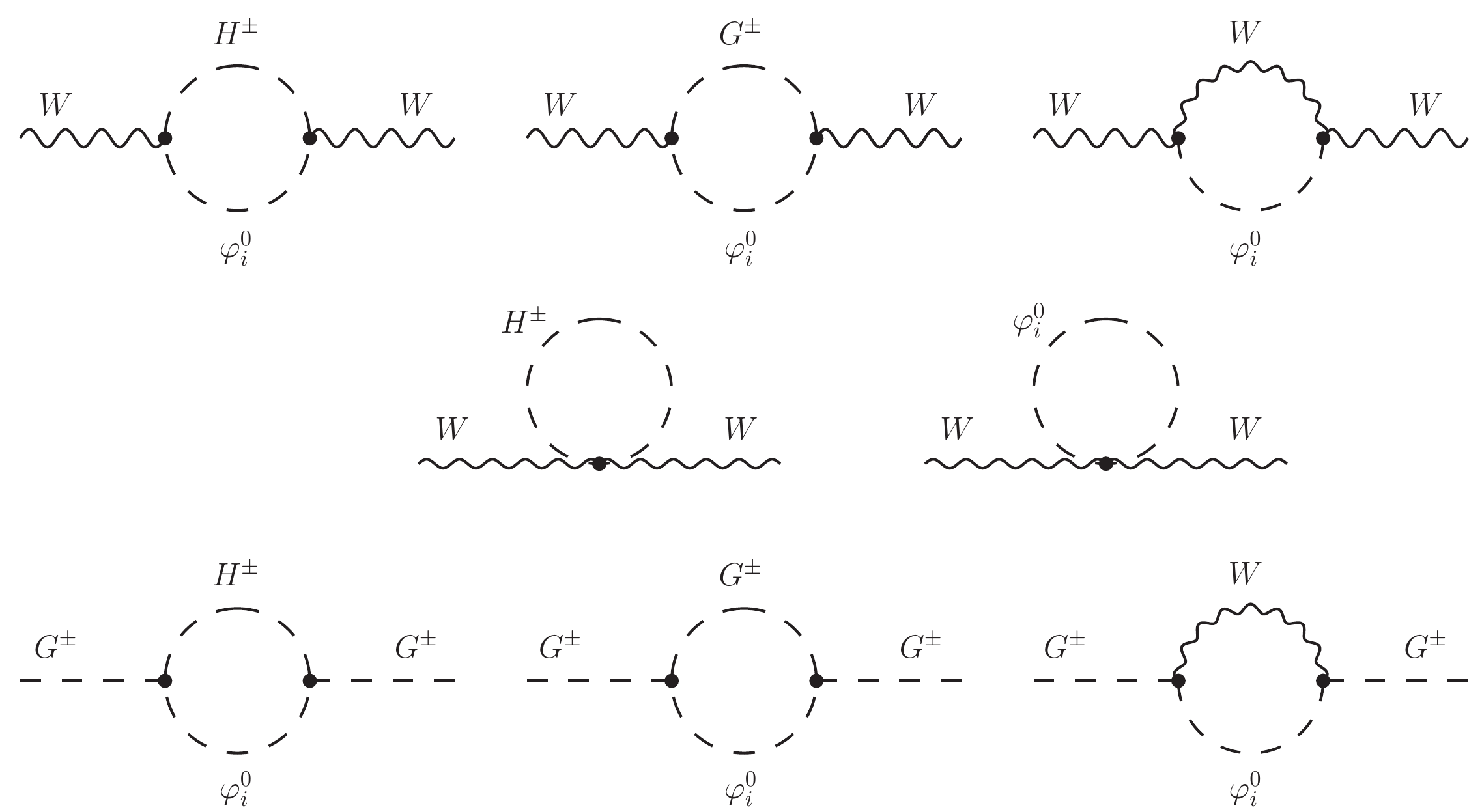}
\caption{{\it One-loop $W$ self energy diagrams needed for the vector boson wave-function and mass renormalization, and $G^\pm$ self-energy diagrams needed for the charged Goldstone boson wave-function renormalization.  }}
\label{WGSelfEnergy}
\end{figure}

The one-loop counterterms for the needed $WW\gamma$ and $WG\gamma$ vertices are given by
\begin{align}
 i \; \Gamma^{\rho\mu\nu}_\delta = i \; e \;  \Gamma^{\rho\mu\nu} \; \delta_W \; , &&
 i \; \Gamma^{\mu\nu}_\delta = i \; e \;  g^{\mu\nu} \; \frac{1}{2}\; (\delta_W + \delta_{G^\pm}+ \delta_M) \, ,
\end{align} 
where $i \,  e  \, \Gamma^{\rho\mu\nu}$ is the tree-level $WW\gamma$ vertex and where we have defined the $G^\pm$, $W^\mu$ and $M_W^2$ renormalization constants as
\begin{align}
Z_W = 1+\delta_W \; , && Z_{G^\pm}=1+\delta_{G^\pm} \; , && Z_M = 1+\delta_M \; .
\end{align}
The needed $W$ and $G^\pm$ self-energy diagrams needed for the calculation of these counter-terms are shown in Fig.~\ref{WGSelfEnergy}.
As we can see, no tadpole diagrams are present. At one-loop level, using the renormalization prescription from \cite{Santos:1996vt}, tadpole
diagrams do not contribute to the $W$ mass renormalization. On the other hand, they do not contribute to the wave-function renormalization either as they do not generate any four-momentum dependence. Thus, for our present calculation we need not to worry
about tadpoles.

\begin{figure}[tbp]
\centering
\includegraphics[scale=0.44]{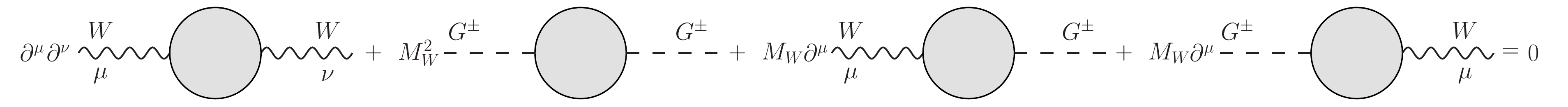}
\caption{{\it One-loop doubly contracted Ward identity. The grey circles stand for the one-loop self energies.}}
\label{WardId}
\end{figure}

One last technical issue is the $W-G^\pm$ mixing that occurs at one-loop level. The gauge fixing Lagrangian cancels exactly the tree-level mixing between the gauge and Goldstone bosons generated by the covariant derivatives. This mixed term, when renormalizing the Lagrangian is in fact, counter-term for the $W-G^\pm$ self-energies, as it is nicely explained in \cite{Santos:1996vt}.  For this calculation, however, we don't need to worry about this mixture. As we are going to ignore the 
propagator corrections, and these corrections are related to the $W-G^\pm$ mixing through the Ward identities (for example the doubly contacted identity shown diagrammatically in 
Fig.~\ref{WardId}), we are also going to ignore the one-loop mixing in order to preserve these identities.

Using the on-shell scheme, working in $D=4+2\epsilon$ dimensions ($\epsilon < 0$), the expression for $\delta_W$ reads $\delta_W=\delta_W^{(1)}+\delta_W^{(2)}+\delta_W^{(3)}$, with:
\begin{align}
\delta_W^{(1)}  \; &=  \; \frac{M_W^2}{v^2}  \; \sum_i |\mathcal{R}_{i2} + \mathcal{R}_{i3}|^2  \; \frac{\mu^{2\epsilon}}{(4\pi)^2}
 \; \Big( \; \frac{1}{3\hat{\epsilon}}  + \int_0^1 dx \;  2x(1-x) \ln \frac{a^2(M_W^2)}{\mu^2}  \; \Big) \; , \\[2ex]
\delta_W^{(2)}  \; &=  \; \frac{M_W^2}{v^2}  \; \sum_i \mathcal{R}_{i1}^2  \; \frac{\mu^{2\epsilon}}{(4\pi)^2}
 \; \Big( \; \frac{1}{3\hat{\epsilon}}  + \int_0^1 dx \;  2x(1-x) \ln \frac{\bar{a}^2(M_W^2)}{\mu^2}  \; \Big) \; , \\[2ex]
\delta_W^{(3)}  \; &=  \; - \frac{4 M_W^4}{v^2}  \; \sum_i \mathcal{R}_{i1}^2  \; \frac{1}{(4\pi)^2}
 \; \int_0^1 dx \;  \frac{x(1-x)}{\bar{a}^2(M_W^2)} \; .
\end{align}
in agreement with \cite{Malinsky:2003bd}. The wave function renormalization counter-term for the charged Goldstone boson is given by $\delta_{G^\pm}=\delta_{G^\pm}^{(1)}+\delta_{G^\pm}^{(1)}+\delta_{G^\pm}^{(1)}$ with:
\begin{align}
\delta_{G^\pm}^{(1)}  \; &=  \; - \frac{1}{(4\pi)^2}  \; \sum_i |\mathcal{R}_{i2} + \mathcal{R}_{i3}|^2  \;
 \frac{(M_{\varphi_i^0}^2-M_{H^\pm}^2)^2}{v^2} \; \int_0^1 dx \;  \frac{x(1-x)}{a^2(M_W^2)} \; , 
\\[2ex]
 \delta_{G^\pm}^{(2)}  \; &=  \; - \frac{1}{(4\pi)^2}  \; \sum_i \mathcal{R}_{i1}^2 \;
 \frac{M_{\varphi_i^0}^4}{v^2} \; \int_0^1 dx \;  \frac{x(1-x)}{\bar{a}^2(M_W^2)} \; , 
\\[2ex]
\delta_{G^\pm}^{(3)}  \; &=  \; - \frac{\mu^{2\epsilon}}{(4\pi)^2} \; \frac{M_W^2}{v^2} \; \sum_i \mathcal{R}_{i1}^2 \;
 \Big[ \; \frac{2}{\hat{\epsilon}} + \frac{1}{6}  + \int_0^1 dx \; \big( 3x^2 - 6x + 4 \big) \ln \frac{\bar{a}^2(M_W^2)}{\mu^2} \; 
 \notag \\
 & \qquad\qquad\qquad + \; \int_0^1 dx \; \frac{x(x-1)}{\bar{a}^2(M_W^2)} \big( M_W^2 (3x^2-8x+6)+2xM_{\varphi_i^0}^2 \big)
 \; \Big] \; .
\end{align}
Last, the $W$ mass counter-term is given by $\delta_M = \delta_M^{(1)}+\delta_M^{(2)}+\delta_M^{(3)}+\delta_M^{(4)}+\delta_M^{(5)}$ with:

\begin{align}
\delta_{M}^{(1)}  \; &=  \; \frac{1}{v^2} \; \frac{\mu^{2\epsilon}}{(4\pi)^2}  \; \sum_i |\mathcal{R}_{i2} + \mathcal{R}_{i3}|^2  \;
\Big[ \;  \Big (\frac{1}{\hat{\epsilon}}-1 \Big)\big( M_{H^\pm}^2 + M_{\varphi_i^0}^2 - \frac{1}{3}M_W^2 \big) \notag 	\\
 & \qquad\qquad\qquad\qquad\qquad\qquad\qquad\qquad + \; \int_0^1 dx \;  2a^2(M_W^2) \ln \frac{a^2(M_W^2)}{\mu^2}   \;  \Big] 
  \; , \\[2ex]
\delta_{M}^{(2)}  \; &=  \; \frac{1}{v^2} \; \frac{\mu^{2\epsilon}}{(4\pi)^2}  \; \sum_i {R}_{i1}^2  \;
\Big[ \;  \Big (\frac{1}{\hat{\epsilon}}-1 \Big)\big( M_{\varphi_i^0}^2 + \frac{2}{3}M_W^2  \big) + \; \int_0^1 dx \;  2\bar a^2(M_W^2) \ln \frac{\bar a^2(M_W^2)}{\mu^2}   \;  \Big] 
  \; ,
\\[2ex]
\delta_{M}^{(3)}  \; &=  \; -\frac{4M_W^2}{v^2} \; \frac{\mu^{2\epsilon}}{(4\pi)^2}  \; \sum_i {R}_{i1}^2  \;
\Big[ \;  \frac{1}{\hat{\epsilon}} + \; \int_0^1 dx \;  \ln \frac{\bar a^2(M_W^2)}{\mu^2}   \;  \Big] 
  \; ,
\\[2ex]
\delta_{M}^{(4)}  \; &=  \; -\frac{2M_{H^\pm}^2}{v^2} \; \frac{\mu^{2\epsilon}}{(4\pi)^2}  \;
\Big[ \;  \frac{1}{\hat{\epsilon}} + \;  \ln \frac{M_{H^\pm}^2}{\mu^2} -1  \;  \Big] 
  \; , \\[2ex]
\delta_{M}^{(5)}  \; &=  \; - \frac{\mu^{2\epsilon}}{(4\pi)^2}  \; \sum_i \frac{M_{\varphi_i^0}^2}{v^2} \;
\Big[ \;  \frac{1}{\hat{\epsilon}} + \;  \ln \frac{M_{\varphi_i^0}^2}{\mu^2} -1  \;  \Big] 
  \; .
\end{align}
Here we have defined $1/\hat{\epsilon} \equiv 1/\epsilon + \gamma_E - \ln(4\pi)$. The functions $a^2(p^2)$ and $\bar{a}^2(p^2)$ are given by:
\begin{align}
a^2(p^2) \; &= \; -p^2 \; x(1-x) + M_{\varphi_i^0}^2 \; x \; + M_{H^\pm}^2 \; (1-x) \; , \\
\bar{a}^2(p^2) \; &= \; -p^2 \; x(1-x) + M_{\varphi_i^0}^2 \; x \; + M_W^2 \; (1-x) \; .
\end{align}

Now, we move on and present the expressions for the one-loop $WW\gamma$ effective vertices from Fig.~\ref{WWgeffV}. The considered kinematics and the assigned Lorentz indices for this process are $W^+(k-q,\rho)+\gamma(q,\mu)\to W^+(k,\nu)$.  Discarding all terms proportional to $q^\mu$, the first and second diagrams give:
\begin{align}
i \; \Gamma^{\rho\mu\nu}_{(1)} &= i \frac{e}{(4\pi)^2} \frac{M_W^2}{v^2} \mu^{2\epsilon} \sum_i |\mathcal{R}_{i2}+\mathcal{R}_{i3}|^2
\Big[ -\frac{1}{3\hat{\epsilon}} \Gamma^{\rho\mu\nu} + \int_0^1 dx \int_0^1 dy \, J^{\rho\mu\nu}_{(a)} \; 2(1-x) \ln
\frac{a_{\text{x}}^2}{\mu^2} \notag \\
& \qquad\qquad\qquad\qquad\qquad\qquad\qquad\qquad + \; \int_0^1 dx \int_0^1 dy \, \frac{J^{\rho\mu\nu}_{(b)}}{k^2-M_{\text{x}}^2 - 2 y \; k\cdot q} \;
 \Big] \; ,
\\[2ex]
i \; \Gamma^{\rho\mu\nu}_{(2)} &= i \frac{e}{(4\pi)^2} \frac{M_W^2}{v^2} \mu^{2\epsilon} \sum_i \mathcal{R}_{i1}^2
\Big[ -\frac{1}{3\hat{\epsilon}} \Gamma^{\rho\mu\nu} + \int_0^1 dx \int_0^1 dy \, J^{\rho\mu\nu}_{(a)} \; 2(1-x) \ln
\frac{\bar{a}_{\text{x}}^2}{\mu^2} \notag \\
& \qquad\qquad\qquad\qquad\qquad\qquad\qquad\qquad + \; \int_0^1 dx \int_0^1 dy \, \frac{J^{\rho\mu\nu}_{(b)}}{k^2-\bar{M}_{\text{x}}^2 - 2 y \; k\cdot q} \; 
 \Big] \; .
\end{align}
Again, $\Gamma^{\rho\mu\nu}$ is the tree-level vertex function and it is given by
\begin{align}
\Gamma^{\rho\mu\nu} = g^{\mu\nu} (-k-q)^\rho + g^{\mu\rho}(2q-k)^\nu + g^{\nu\rho}(2k-q)^\mu \, .
\end{align}
The sum of diagrams (3), (4) and (5) gives
\begin{align}
i \; \Gamma^{\rho\mu\nu}_{(3+4+5)} &= - i \frac{e}{(4\pi)^2} \frac{M_W^4}{v^2} \sum_i \mathcal{R}_{i1}^2
 \; \int_0^1 dx \int_0^1 dy \; \frac{1}{x} \; \frac{-2 \, g^{\mu\nu} Q^\rho  - 2 \, g^{\mu\rho}J^\nu + 4 \, J^{\rho\mu\nu}_{(c)}}{k^2-\bar{M}_{\text{x}}^2 - 2 y \; k\cdot q} \;
  \; .
\end{align}
With diagram (6) we have to be specially careful. Its explicit expression reads
\begin{align}
i \; \Gamma^{\rho\mu\nu}_{(6)} &=  i \frac{e}{(4\pi)^2} \frac{M_W^2}{v^2} \, g^{\mu\nu}(k^\rho - q^\rho) \; \mu^{2\epsilon} \, 
\sum_i |\mathcal{R}_{i2} + i \mathcal{R}_{i3} |^2
 \; \int_0^1 dx \; (2x-1) \, \Big(\frac{1}{\hat{\epsilon}} + \ln \frac{b^2_{\text{x}}}{\mu^2}\Big) \; .
\end{align}
Integrating over $x$, the pole and the $\mu$-dependence vanish. We are left with a logarithm that depends on the four momentum and that we need to integrate in the second loop. Using the expansion ($\delta\ll 1$)
\begin{align}
\ln A = \frac{1}{\delta} \Big( A^{\delta} -1 \Big) + O(\delta) \; ,
\end{align}
we can write the previous expression as
\begin{align}
i \; \Gamma^{\rho\mu\nu}_{(6)} &= i \frac{e}{(4\pi)^2} \frac{M_W^2}{v^2} \, g^{\mu\nu}(k^\rho - q^\rho) \, \frac{(-1)^\delta}{\delta}  \, 
\sum_i |\mathcal{R}_{i2} + i \mathcal{R}_{i3} |^2
 \; \int_0^1 dx \; \frac{(2x-1) \, x^\delta (1-x)^\delta}{(k^2-M_{\text{x}}^2-2 \, k\cdot q)^{-\delta}}  \; ,
\end{align}
and use the Feynman parametrization 
\begin{align}
\frac{1}{A_1^{-\delta}A_2A_3A_4} &= \frac{\Gamma(3-\delta)}{\Gamma(-\delta)}\int_0^1 dx_1 \int_0^{1-x_1} dx_2
\int_0^{1-x_1-x_2} dx_3 \;\; x_1^{-\delta-1}  \notag \\
 & \qquad \qquad \qquad \times \; \frac{ 1 }{\big(x_1 A_1 + x_2 A_2 + x_3 A_3 + (1-x_1-x_2-x_3) A_4 \big)^{3-\delta}} \; ,
\end{align}
in order to solve the second loop (taking the limit $\delta\to 0$ at the end of the calculation). We obtain a similar expression 
for diagram (7):
\begin{align}
i \; \Gamma^{\rho\mu\nu}_{(7)} &= i \frac{e}{(4\pi)^2} \frac{M_W^2}{v^2} \, g^{\mu\nu}(k^\rho - q^\rho) \, \frac{(-1)^\delta}{\delta}  \, 
\sum_i \mathcal{R}_{i1}^2 
 \; \int_0^1 dx \; \frac{(2x-1) \, x^\delta (1-x)^\delta}{(k^2-\bar{M}_{\text{x}}^2-2 \, k\cdot q)^{-\delta}}  \, .
\end{align}
Contributions (8) and (9) vanish as their expressions are terms proportional to $q^\mu$. Finally, diagrams (10) and (11) read
\begin{align}
i \; \Gamma^{\rho\mu\nu}_{(10)} &= i \frac{e}{(4\pi)^2} \frac{M_W^2}{v^2} \, g^{\mu\rho} k^\nu \, \mu^{2\epsilon} \,  
\sum_i |\mathcal{R}_{i2} + i \mathcal{R}_{i3} |^2
 \; \int_0^1 dx \; (2x-1) \,   \Big(\frac{1}{\hat{\epsilon}} + \ln \frac{c^2_{\text{x}}}{\mu^2}\Big) \; , \\
i \; \Gamma^{\rho\mu\nu}_{(11)} &= i \frac{e}{(4\pi)^2} \frac{M_W^2}{v^2} \, g^{\mu\rho} k^\nu \, \mu^{2\epsilon} \,  
\sum_i \mathcal{R}_{i1}^2 
 \; \int_0^1 dx \; (2x-1) \,   \Big(\frac{1}{\hat{\epsilon}} + \ln \frac{\bar{c}^2_{\text{x}}}{\mu^2}\Big) \; , 
\end{align}
which can be treated exactly as diagrams (6) and (7). The previously introduced tensorial functions are given by:
\begin{align}
J^{\rho\mu\nu}_{(a)} &= g^{\mu\rho}\big( (1-2x) k^\nu + 2y(x-1)q^\nu \big) + g^{\mu\nu} \big( (1-2x)k^\rho + 
(2(x-1)y+1)q^\rho \big) - 2x \, g^{\nu\rho} \, k^\mu \; , \\
J^{\rho\mu\nu}_{(b)} &= -2 k^\mu \big( (2x-1)k^\nu - 2y(x-1)q^\nu  \big)\big( (1-2x)k^\rho + (2(x-1)y+1)q^\rho \big)  \; ,  \\
J^{\rho\mu\nu}_{(c)} &= g^{\mu\rho}\big( ( xy - y + 2  ) q^\nu - x k^\nu \big) - g^{\mu\nu} \big( x k^\rho + 
q^\rho(y-xy+1) \big) + 2x \, g^{\nu\rho} \, k^\mu  \; ,
\end{align}
and,
\begin{align}
Q^\rho &= k^\rho \, (1-2x) + q^\rho \, (2xy -2y + 1)  \; , &&&
J^\nu & = k^\nu \, (1-2x) + q^\nu \, 2y(x-1) \; .
\end{align}
The scalar functions are given by:
\begin{align}
a_{\text{x}}^2 &= -x(1-x) (k^2 - M_{\text{x}}^2 - 2 y \; k\cdot q) \; ,  &&&
\bar{a}_{\text{x}}^2 &= -x(1-x) (k^2 - \bar{M}_{\text{x}}^2 - 2 y \; k\cdot q) \; , \notag \\
b_{\text{x}}^2 &= -x(1-x) (k^2 - M_{\text{x}}^2 - 2 \; k\cdot q) \; ,  &&&
\bar{b}_{\text{x}}^2 &= -x(1-x) (k^2 - \bar{M}_{\text{x}}^2 - 2 \; k\cdot q) \; , \\
c_{\text{x}}^2 &= -x(1-x) (k^2 - M_{\text{x}}^2) \; , &&&
\bar{c}_{\text{x}}^2 &= -x(1-x) (k^2 - \bar{M}_{\text{x}}^2) \; ,  \notag
\end{align}
with
\begin{align}
&& M_{\text{x}}^2 = \frac{M_{\varphi_i^0}^2}{1-x} + \frac{M_{H^\pm}^2}{x} \; , &&
\bar{M}_{\text{x}}^2 = \frac{M_{\varphi_i^0}^2}{1-x} + \frac{M_{W}^2}{x} \; . && 
\end{align}
Next we present the $G^\pm W \gamma$ effective vertices from Fig.~\ref{WGgeffV}. The kinematics and Lorentz indices are given by 
$G^+(k-q)+\gamma(q,\mu)\to W^+(k,\nu)$. Thus, the one-loop expressions are:
\begin{align}
i \; \Gamma^{\mu\nu}_{(1)} & = -i \frac{e}{(4\pi)^2} M_W \, \mu^{2\epsilon} \, \sum_i |\mathcal{R}_{i2} + i \mathcal{R}_{i3} |^2
\, \frac{M_{\varphi_i^0}^2-M_{H^\pm}^2}{v^2} \, \Big[ \, g^{\mu\nu} \, \frac{1}{\hat{\epsilon}}  \; + \notag
\\ & \qquad\qquad + \; \int_0^1 dx \int_0^1 dy \; \Big( \; 2 g^{\mu\nu} \, (1-x)\ln\frac{a_{\text{x}}^2}{\mu^2}      
- \frac{2 \, K^{\mu\nu}}{k^2-M_{\text{x}}^2 - 2y\, k\cdot q} \, \Big)
  \; \Big] \, , 
\\[2ex]
i \; \Gamma^{\mu\nu}_{(2)} & = -i \frac{e}{(4\pi)^2} M_W \, \mu^{2\epsilon} \, \sum_i \mathcal{R}_{i1}^2 
\, \frac{M_{\varphi_i^0}^2}{v^2} \, \Big[ \, g^{\mu\nu} \, \frac{1}{\hat{\epsilon}}  \; + \notag
\\ & \qquad\qquad + \; \int_0^1 dx \int_0^1 dy \; \Big( \; 2 g^{\mu\nu} \, (1-x)\ln\frac{\bar{a}_{\text{x}}^2}{\mu^2}      
- \frac{2 \, K^{\mu\nu}}{k^2-\bar{M}_{\text{x}}^2 - 2y\, k\cdot q} \, \Big)
  \; \Big] \, ,\\[2ex]
i \; \Gamma^{\mu\nu}_{(3)} & = -i \frac{e}{(4\pi)^2} \, M_W \, \sum_i \mathcal{R}_{i1}^2 
\, \frac{M_{\varphi_i^0}^2}{v^2} \, \int_0^1 dx \int_0^1 dy \; \frac{1}{x}  \;  \frac{2 \, M_W^2 \; g^{\mu\nu}}{k^2-\bar{M}_{\text{x}}^2 - 2y\, k\cdot q} \, ,
\\[2ex]
i \; \Gamma^{\mu\nu}_{(4)} & = i \frac{e}{(4\pi)^2} \frac{M_W^3}{v^2} \, \mu^{2\epsilon} \, \sum_i \mathcal{R}_{i1}^2 
\,  \Big[ \, g^{\mu\nu} \, \frac{1}{2\hat{\epsilon}}  + \int_0^1 dx \int_0^1 dy \; \Big( \; g^{\mu\nu} \, (1-x)\ln\frac{\bar{a}_{\text{x}}^2}{\mu^2} \, + \notag
\\ & \qquad\qquad\qquad\qquad\qquad\qquad\qquad + \;         
 \frac{(2-x)}{x} \; \frac{ K^{\mu\nu}}{k^2-\bar{M}_{\text{x}}^2 - 2y\, k\cdot q} \, \Big)
  \; \Big] \, , \\[2ex]
i \; \Gamma^{\mu\nu}_{(5)} & = i \frac{e}{(4\pi)^2} \frac{M_W^3}{v^2} \, \mu^{2\epsilon} \, \sum_i \mathcal{R}_{i1}^2 
\,  \Big[ - g^{\mu\nu} \, \Big( \frac{3}{2\hat{\epsilon}} + 1 \Big) + \int_0^1 dx \int_0^1 dy \; \Big( \; 3 g^{\mu\nu} \, (x-1)\ln\frac{\bar{a}_{\text{x}}^2}{\mu^2} \, + \notag
\\ & \qquad\qquad\qquad\qquad\qquad\qquad\qquad + \;         
 \frac{2}{x} \; \frac{ G^{\mu\nu}}{k^2-\bar{M}_{\text{x}}^2 - 2y\, k\cdot q} \, \Big)
  \; \Big] \, ,\\[2ex]
i \; \Gamma^{\mu\nu}_{(6)} & = i \frac{e}{(4\pi)^2} M_W \, g^{\mu\nu} \, \mu^{2\epsilon} \, \sum_i \mathcal{R}_{i1}^2 
\, \frac{M_{\varphi_i^0}^2}{v^2} \, \Big( \, \frac{1}{\hat{\epsilon}}  \; + 
 \int_0^1 dx \; \ln\frac{\bar{b}_{\text{x}}^2}{\mu^2} \, \Big) \, ,\\[2ex]
i \; \Gamma^{\mu\nu}_{(7)} & = i \frac{e}{(4\pi)^2} M_W \, g^{\mu\nu} \, \mu^{2\epsilon} \, \sum_i |\mathcal{R}_{i2} + \mathcal{R}_{i3}|^2 
\, \frac{M_{\varphi_i^0}^2 - M_{H^\pm}^2}{v^2} \, \Big( \, \frac{1}{\hat{\epsilon}}  \; + 
 \int_0^1 dx \; \ln\frac{b_{\text{x}}^2}{\mu^2} \, \Big) \, ,
\\[2ex]
i \; \Gamma^{\mu\nu}_{(8)} & = i \frac{e}{(4\pi)^2} \, g^{\mu\nu} \,
\frac{2 M_W^3}{v^2} \, \mu^{2\epsilon} \,  \sum_i \mathcal{R}_{i1}^2  
  \, \Big( \, \frac{1}{\hat{\epsilon}}  \; + \; \int_0^1 dx \; 
  \ln\frac{\bar{c}_{\text{x}}^2}{\mu^2} \, \Big) \, .  
\end{align}
The tensorial functions are given by:
\begin{align}
K^{\mu\nu} &= k^\mu \; \big( (2x-1) \, k^\nu - 2y (x-1) \,  q^\nu \big) \; , \\
G^{\mu\nu} &= g^{\mu\nu}\big(k^2 x(x-2) - 2(x-1)(xy-y-1) k\cdot q \big) + k^\mu \big( (x-1)(xy+2y-4) q^\nu - 
x(x-2) k^\nu \big) .
\end{align}
All other functions are the same as previously. Note, that for the previous expressions of the one-loop effective vertices, we have maintained the $k \cdot q$ structure in the denominator (in contrast to the $H^\pm W\gamma$ effective vertices) because here, in some cases, this structure does contribute to the final result.

Inserting all these expressions into the second loop we finally obtain the expression
for the total contribution to the anomalous magnetic moment of the muon. Subtracting the SM contributions we have  
\begin{align}
\Delta a_\mu \; = \; \frac{\alpha}{128 \; \pi^2 \; s_{\text{w}}^2} \; \frac{m_\mu^2}{v^2} \;
\int_0^1 dx  \; \Big(	\; \sum_i  \mathcal{R}_{i1}^2 \, \mathcal{A} - \mathcal{A}_{\text{SM}} \, + \sum_i  |\mathcal{R}_{i2}+i\mathcal{R}_{i3}|^2 \,
\mathcal{B} + \mathcal{C} \; \Big) \; ,
\end{align}
with the functions $\mathcal{A}$, $\mathcal{B}$ and $\mathcal{C}$ given by:
\begin{align}
\mathcal{A}  = \; & \frac{7}{3} x(1-x) \ln\frac{\bar{a}^2(M_W^2)}{M_W^2} -
\frac{(2x^2-3x+2)\;M_{\varphi_i^0}^2}{2x (M_W^2 - \bar{M}_{\text{x}}^2)}
   + \frac{6(x-1)\; \bar{M}_{\text{x}}^2 + (-12x^2+30x-55)\, M_W^2}{6 (M_W^2-\bar{M}_{\text{x}}^2)} \, + \notag \\[2ex]
& \, + \, 
  \frac{  M_{\varphi_i^0}^2 \; \ln(\bar{M}_{\text{x}}^2/M_W^2)}{2x M_W^2 (M_W^2-\bar{M}_{\text{x}}^2)^2} \, 
  \Big( \bar{M}_{\text{x}}^4 \; x(2x-1) - 2 M_W^4 + 4 M_W^2 \bar{M}_{\text{x}}^2 \; x(1-x)   \Big) \; + \notag
\\[2ex]
& \, + \, 
\frac{\ln( \bar{M}_{\text{x}}^2/M_W^2 )}{6x \, (M_W^2-\bar{M}_{\text{x}}^2)^2} \, 
  \Big( \bar{M}_{\text{x}}^4 \; x(16x-9) +  M_W^4 \, (8x-42) + 2 M_W^2 \bar{M}_{\text{x}}^2 \; (-6x^3+10x^2-30x+21) \Big) \; + \notag
   \\[2ex]
& \, + \, 
\frac{x(1-x)}{4M_W^2 \; \bar{a}^2(M_W^2)} \, 
  \Big(  M_W^4 \, \big(3x^2-8x-\frac{50}{3} \big) + 2x \, M_W^2 M_{\varphi_i^0}^2  \; - M_{\varphi_i^0}^4   \Big) \; , \\[3ex]
\mathcal{B}  = \; & \frac{7}{3} x(1-x) \ln\frac{a^2(M_W^2)}{M_W^2} + \frac{1}{2}(2x-1) \frac{M_\text{x}^2-2M_W^2(x-1)}{M_W^2-M_{\text{x}}^2} + \frac{(M_{\varphi_i^0}^2-M_{H^\pm}^2)(3-2x)}{2(M_W^2-M_\text{x}^2)}  + \notag \\[2ex]
& \, + \, \frac{M_\text{x}^2 \ln (M_\text{x}^2/M_W^2)}{6(M_W^2-M_\text{x}^2)^2}
\Big(  M_W^2 2x(7-6x) + M_\text{x}^2 (10x-9)   \Big) - \frac{(M_{\varphi_i^0}^2-M_{H^\pm}^2)^2 \, x(1-x)}{4M_W^2 \; a^2(M_W^2)}
+ \notag \\[2ex]
& \, + \, \frac{(M_{\varphi_i^0}^2-M_{H^\pm}^2) \, M_\text{x}^2 \ln(M_\text{x}^2/M_W^2) }{2M_W^2(M_W^2-M_\text{x}^2)^2} \;
\Big(      4M_W^2 (1-x) + M_\text{x}^2 (2x-1)            \Big) \;  + \notag \\[2ex]
 & \, + \, \frac{1}{4 M_W^2} \, \Big( 
2(1-x) \; M_{H^\pm}^2 \ln \frac{a^2(M_W^2)}{M_{H^\pm}^2}  + 2x \; M_{\varphi_i^0}^2
\ln \frac{a^2(M_W^2)}{M_W^2}
  \; \Big) \; ,
\\[3ex]
  \mathcal{C}  = & \sum_i \Big(  
 -\frac{M_{\varphi_i^0}^2}{4M_W^2}\ln\frac{M_{\varphi_i^0}^2}{M_W^2} \; + \mathcal{R}_{i1}^2 \, \frac{1}{4}(-3 x^2 + 4x - 6) \ln\frac{\bar{a}^2(M_W^2)}{\bar{a}^2_{\text{SM}}(M_W^2)}   \; + \notag \\[2ex]
   & \qquad\qquad\qquad\qquad +  \,  \mathcal{R}_{i1}^2 \, \frac{x\, M_{\varphi_i^0}^2}{2M_W^2}\ln\frac{\bar{a}^2(M_W^2)}{M_W^2}   \Big)
      - 
  \frac{x\, M_\phi^2}{2M_W^2}\ln\frac{\bar{a}^2_{\text{SM}}(M_W^2)}{M_\phi^2} +\frac{1}{6}\; .
\end{align}
All the functions that carry a SM sub-index are obtained from the original ones by replacing $M_{\varphi_i^0}$ with $M_\phi$ everywhere, where $M_\phi$ is the mass of the SM Higgs. The numerical values that we obtain for this contribution (for $M_{H,A,H^\pm} < 500$ GeV) are typically of $\mathcal{O}(10^{-11})$ both positive or negative, which is two orders of magnitude below $\Delta a_\mu^{exp}$, therefore we shall not take it into account in this analysis.

\end{appendix}

%%%%%%%%%%%%%%%%%%%%%%%%%%%%%%%%%%%%%%%%%%%%%%%%%%%%%%%%%%%%%%%%%%%%%%%%%%%%%%%%%%%%%%%%%%%%%%%%%%%%%%%%%%%%%%%%%%%%%%%%%%
%%%%%%%%%%%%%%%%%%%%%%%%%%%%%%%%%%%%%%%%%%%%%%%%%%%%%%%%%%%%%%%%%%%%%%%%%%%%%%%%%%%%%%%%%%%%%%%%%%%%%%%%%%%%%%%%%%%%%%%%%%
%%%%%%%%%%%%%%%%%%%%%%%%%%%%%%%%%%%%%%%%%%%%%%%%%%%%%%%%%%%%%%%%%%%%%%%%%%%%%%%%%%%%%%%%%%%%%%%%%%%%%%%%%%%%%%%%%%%%%%%%%%

\section*{Acknowledgements}
I am grateful to Antonio Pich for reviewing this manuscript and to Alejandro Celis for helpful comments. This work has been supported in part by the Spanish Government and ERDF funds from the
EU Commission [Grants FPA2011-23778 and CSD2007-00042 (Consolider Project CPAN)] and
by the Spanish Ministry MINECO through the FPI grant BES-2012-054676.

\end{document}